\documentclass[pra,aps,twocolumn,superscriptaddress,notitlepage,nofootinbib,floatfix]{revtex4-2}
\usepackage[utf8]{inputenc}
\usepackage{bm}
\usepackage{bbm}
\usepackage{epsfig}
\usepackage{wrapfig}
\usepackage{float}
\usepackage{mathrsfs}
\usepackage{amsmath,amssymb,amsfonts}
\usepackage{graphicx,xcolor}
\usepackage{comment}

\usepackage[bookmarks=false, pdfstartview=FitH]{hyperref}

\newcommand{\discretep}[2]{\ensuremath{P_{\{#1\}}(#2)}}

\begin{document}

\title{The Grasshopper Problem on the Sphere}

\author{David Llamas}
\affiliation{Department of Physics, University of Massachusetts Boston, Boston, MA 02125, USA}

\author{Dmitry Chistikov}
\affiliation{Centre for Discrete Mathematics and its Applications (DIMAP) \& Department of Computer Science, University of Warwick,
Coventry, CV4 7AL, U.K.}

\author{Adrian Kent}
\email{apak@cam.ac.uk}
\affiliation{Centre for Mathematical Sciences, University of Cambridge, Wilberforce Road, Cambridge, CB3 0WA, United Kingdom.}
\affiliation{Perimeter Institute for Theoretical Physics, 31 Caroline Street North, Waterloo, Ontario N2L 2Y5, Canada.}

\author{Mike Paterson}
\affiliation{Centre for Discrete Mathematics and its Applications (DIMAP) \& Department of Computer Science, University of Warwick,
Coventry, CV4 7AL, U.K.}

\author{Olga Goulko}
\email{olga.goulko@umb.edu}
\affiliation{Department of Physics, University of Massachusetts Boston, Boston Massachusetts 02125, USA}

\date{\today}

\begin{abstract}
The spherical grasshopper problem is a geometric optimization problem that arises in the context of Bell inequalities and can be interpreted as identifying the best local hidden variable approximation to quantum singlet correlations for measurements along random axes separated by a fixed angle.
In a parallel publication \cite{llamas2025sphericalletter}, we presented numerical solutions for this problem and explained their significance for singlet simulation and testing.
In this companion paper, we describe in detail the geometric and computational framework underlying these results. 
We examine the role of spherical discretization and compare three natural variants of the problem: antipodal complementary lawns, antipodal independent lawns, and non-antipodal complementary lawns.
We analyze the geometric structure of the corresponding optimal lawn configurations and interpret it in terms of a spherical harmonics expansion. We also discuss connections to other physical models and to classical problems in geometric probability.
\end{abstract}

\maketitle

\section{Introduction}
\label{sec:introduction}
A simple version of the grasshopper problem on the sphere is formulated as follows. Half of the area of a unit sphere is covered by a lawn, such that exactly one of every pair of antipodal points belongs to the lawn. A grasshopper lands at a uniformly chosen random point on the lawn, and then jumps in a random direction through a spherical angle $\theta$. What lawn shape maximizes the probability that the grasshopper remains on the lawn after jumping, and what is this maximum probability as a function of $\theta$?

The original motivation for studying this problem and the extensions considered in the following is to synthesize and extend the known results on Bell inequalities.
These distinguish the correlations predicted by quantum theory for bipartite systems from the possible correlations that can be predicted by any locally causal theory.   In principle, this gives a simple way of designing experiments that can confirm quantum theory and refute all locally causal theories.   In practice, inferring the refutation of locally causal theories requires assumptions about the experiment, leading to logical loopholes in this clean interpretation. Although the most commonly cited loopholes have now been closed \cite{HBDRKBRVSAAPMMTEWTH15,SMCBWSGGHACDHLVLTMZSAAPJMKBMKN15,GVMWHHPSKLAAPMBGLSNSUWZ15}, some interesting possibilities still remain (e.g. \cite{K05,K18.2.1,K20}). However, the general consensus is that we have compelling evidence that nature is not locally causal.  Our discussion takes as given that quantum theory is correct.   Our aim is to identify new ways in which quantum non-locality can be quantified.   These have practical applications as well as intrinsic theoretical interest, since in many scenarios noise, errors, or adversarial action can replace bipartite states that ideally would be maximally entangled and hence maximally non-local by states that actually could be described by local hidden variables.   Comparing different measures of non-locality allows us to analyze which tests most efficiently identify when this has happened.     

The theory of quantum Bell non-locality began with Bell's proof \cite{Bell} that quantum correlations for projective spin measurements on two spin-$1/2$ particles in a singlet state cannot be reproduced by any model in which the outcomes for all possible spin measurements are determined in advance as separate functions of the separate measurement choices.   This holds for probabilistic as well as deterministic functions.   
A projective spin measurement is defined by a choice of axis ${\mathbf a}$ in three dimensions, and produces outcomes that we label by $\pm 1$.   For axes ${\mathbf a},{\mathbf b}$, the expected value of the outcome product is 
\begin{equation}
    E ( {\mathbf a},{\mathbf b} ) = -{\mathbf a}.{\mathbf b}  \, .  
\end{equation}
Bell's theorem shows that this cannot be reproduced by any expression of the form
\begin{equation}
    \int f_A ( {\mathbf a}, \lambda ) f_B ({\mathbf b}, \lambda ) d \lambda \, , 
\end{equation}
where $f_A , f_B$ take values $\pm 1$ and depend only on the respective axis choices ${\mathbf a},{\mathbf b}$ and a ``hidden variable'' parameter $\lambda$ that is independent of ${\mathbf a},{\mathbf b}$.   
Bell inequalities identify specific combinations of measurement choices (here, axis choices) that distinguish between the predictions of quantum theory for a given quantum state (here, the spin singlet) and the predictions of any possible Bell-local ``hidden variable'' theory.
Clauser et al. (CHSH) \cite{CHSH69} identified a simple Bell inequality that is well-adapted for experimental tests: any Bell-local theory satisfies
\begin{multline}
  S_{\rm CHSH} = \\
  \qquad | E ( {\mathbf a_1},{\mathbf b_1} ) +   E ( {\mathbf a_2},{\mathbf b_1} )  +
    E ( {\mathbf a_2},{\mathbf b_2} ) -    E ( {\mathbf a_1},{\mathbf b_2} ) | \leq 2 
\end{multline} 
for any pairs of possible measurement choices ${\mathbf a_1},{\mathbf a_2}$ and 
${\mathbf b_1},{\mathbf b_2}$.    
In comparison, for specific measurement choices on the singlet state, in which the axis pairs 
are separated by $\pi/4$, quantum theory predicts 
$S_{\rm CHSH} =2 \sqrt{2}>2$. 
Braunstein-Caves (BC) \cite{BC90} noted a useful generalisation with $n$ measurement choices on each system: in any Bell-local theory
\begin{multline}
 S_{\rm BC} = \\
 | E ( {\mathbf a_1},{\mathbf b_1} ) +   E ( {\mathbf a_2},{\mathbf b_1}) + E ( {\mathbf a_2},{\mathbf b_2})  +E ( {\mathbf a_3},{\mathbf b_2})  + \cdots \\
\cdots + E ({\mathbf a_n},{\mathbf b_{n-1}} ) + E ({\mathbf a_n},{\mathbf b_n} ) - E ({\mathbf a_1},{\mathbf b_n} ) |  \leq 2n-2 \, .    
\end{multline}

For specific measurement choices on the quantum singlet state, when each axis pair is separated by $\pi/(2n)$, 
quantum theory predicts $ S_{\rm BC} = 2n \cos (\pi / 2n ) > (2n-2)$.     

In both cases, the maximal expression for a Bell-local theory is attained by Bell's simple 
hemispherical model \cite{Bell}.   This takes the hidden variable to be a random axis ${\mathbf c}$ and
\begin{equation}
f_A ( {\mathbf a}, {\mathbf c} ) =  {\rm sgn} ( {\mathbf a}.{\mathbf c} ) \, , \qquad 
f_B ( {\mathbf b}, {\mathbf c} ) = -{\rm sgn} ( {\mathbf b}.{\mathbf c}) \, , 
\end{equation}
where ${\rm sgn}(x) = 1,-1$ for $x>0,x<0 $ respectively.
For generic ${\mathbf c}$ the functions are defined and $f_A ( {\mathbf a} ) = - f_B ({\mathbf a})$,
reproducing the perfect anti-correlations predicted by quantum theory for the singlet.

(Bell left the model undefined in the cases ${\mathbf a}.{\mathbf c} = 0$ and ${\mathbf b}.{\mathbf c} =0$.
These are measure zero, so do not affect the expectation values.  
We suppose some definitions are chosen that ensure $f_A ( {\mathbf a} ) = - f_B ( {\mathbf a})$ for all ${\mathbf c}$.) 
  
Bell's hemispherical model predicts that each term in the expression for $S_{\rm BC}$, for the given measurement choices, is $-(2n-2)/(2n)$;
for the CHSH case $n=2$ each term is $-1/2$.  
More generally, the model predicts $E({\mathbf a}, {\mathbf b} ) = - (2n-2)/(2n)$ for {\it any} pair of axes ${\mathbf a}, {\mathbf b}$
separated by $\pi/(2n)$, while quantum theory predicts $- \cos (\pi / (2n) )$ for spin measurements about such axes on the singlet. 

Another way of thinking about Bell's original result and the CHSH and BC inequalities is thus that they rely on inequalities between
the maximum anti-correlation obtained by Bell-local hidden variable models and by quantum theory for measurements about randomly chosen
axes separated by a given angle $ \theta$, where $\theta$ is small in Bell's first proof,
$\theta = \pi/4$ in the CHSH case and $\pi/(2n)$ in the BC case. 
This inspires the question, first studied in Ref.~\cite{kent2014bloch}, of what can be said about the gap between the anti-correlations predicted by Bell-local hidden variable models and quantum theory for randomly chosen axes separated by a general fixed angle $\theta$.  
Ref.~\cite{kent2014bloch} showed that there is such a gap for all $\theta$ in the range $0 \leq \theta\leq \pi/3$, and hence an infinite class of new Bell inequalities.   
It also showed that the Bell hemisphere model is not optimal for large values of $\theta$. 
However, the bounds were not shown to be optimal for general $\theta$ and optimal classical models were not identified. It was noted that tighter bounds and classical models closer to optimal would be obtained by solving the version of the grasshopper problem given above. 
The underlying idea is that a lawn defines a deterministic assignment of binary outcomes to measurement directions on the Bloch sphere, i.e.\ a local hidden variable (LHV) model, and the probability that the grasshopper remains on the lawn corresponds to classical correlations between measurements. In this way, the grasshopper problem provides a geometric formulation of the optimal classical approximation to quantum singlet correlations for a fixed angular separation of measurement axes.
It was also noted that optimal bounds and models require solving a version of this problem with two independent overlapping lawns, in which the grasshopper jumps from one lawn and the aim is to maximize the probability that it lands outside the other.   

We find these problems interesting for several reasons.   Singlets are a fundamental resource for quantum communication, cryptography and computing, and it is often important to verify that states distributed between two locations are indeed (at least to very good approximation) singlets.   Particularly in the adversarial context of quantum cryptography, parties also need to verify that they share quantum states rather than systems created by a third party to mimic (as far as possible) the behavior of quantum states while actually behaving predictably (by the third party) under measurements.    The two parties attempting secure quantum cryptography (or communication or computing) will only share a finite number of states, and they want to use as many of these as possible for the intended protocol.   Testing a singlet generally effectively consumes some or all of its entanglement, so one wants tests to be as efficient as possible \cite{cowperthwaite2023comparingsinglettestingschemes}.   This motivates us to explore whether the new Bell inequalities obtainable from solving the grasshopper problem are more efficient than the CHSH, BC or other known tests.   One might think they may be as they involve measurements about general axes in three dimensions, rather than a fixed finite set in a plane: this, one might think, makes it harder for an adversary to pass the tests, since they have much less advance information about the chosen axes.  Also, at first sight, there seems no obvious reason why efficiency should necessarily be maximized in the discrete set of angles required for the CHSH and BC inequalities.  However, these arguments are not conclusive, and (as far as we know) the question can only be resolved by explicitly quantifying the Bell inequalities -- as we do via the numerical investigations below -- and comparing their efficiencies.  

Despite its simple statement, the grasshopper problem (in every version) is analytically challenging and rich in structure. 
Previous work first explored the planar variant of the grasshopper problem using analytical and numerical means \cite{goulko2017grasshopper,llamas2023grasshopper}. It was shown that the disk-shaped lawn is never optimal in two dimensions. Optimal lawn shapes for small jumps exhibit cog-like perturbations, with more complex disconnected lawn shapes emerging for larger jumps. In three and higher dimensions on the other hand, full rotational symmetry of the optimal shapes for small jumps is restored. 
Analytical results on the circle and the sphere \cite{chistikov2020globehopping} also revealed interesting features.  For the sphere, hemispherical lawns were shown to be almost never optimal, implying that optimal LHV models are generically more
complex than Bell's hemispherical model.

Prior numerical results for the spherical grasshopper problem \cite{breugel2023partiii} indicated that for a range of grasshopper jumps optimal grasshopper lawns on the sphere exhibit similar patterns to optimal planar lawns, including cogwheels for small jump distances and stripes for large jumps. Other patterns, such as labyrinth-like configurations, were also observed. These preliminary findings motivated a more detailed systematic numerical study of the problem.
   
In a parallel publication \cite{llamas2025sphericalletter}, we presented more detailed numerical solutions for the spherical grasshopper problem and compared them with the corresponding quantum correlations. We identified numerically optimized lawn configurations across all jump angles $0\leq\theta\leq\pi$ for the single-lawn setup, as well as for a more general setup with two independent lawns, and compared the grasshopper success probabilities with the corresponding quantum singlet probabilities, establishing the maximal size of the gaps between classical and quantum correlations, which yield optimally efficient nonlocality tests in this setup.

The goal of the present work is to provide additional background on and analysis of the numerical methods, including a detailed discussion of different discretization setups, and to give a more comprehensive characterization of the optimal lawn shapes and their geometric properties. To this end we discuss three variants of the spherical grasshopper problem: the antipodal complementary (one-lawn) setup, in which the grasshopper aims to remain on the same antipodal lawn after the jump; the antipodal independent (two-lawn) setup, in which the grasshopper jumps from one lawn and aims to land outside a second lawn; and the non-antipodal complementary (one-lawn) setup, in which the lawn covers half of the area of the sphere but is not restricted to be antipodal.
As reviewed in \cite{llamas2025sphericalletter}, 
the first two define local hidden variable (LHV) models, and optimizing these characterizes how well such models can simulate the anticorrelations of the singlet.
The third is an obviously closely related problem in geometric combinatorics.  It further illustrates our methods and the taxonomy of optimal lawn shapes, and in particular the regimes where the antipodal constraint is or is not significant, but is not directly relevant to simulating quantum correlations by LHVs. 

This paper is structured as follows. Sec.~\ref{sec:problemstatement} presents the formal problem statement for the different setups in the real space and in the spectral representation. Sec.~\ref{sec:method} describes the numerical approach, which is based on a simulated annealing algorithm for the discretized version of the problem, and explains the relevance of the spherical grid resolution and symmetries. Detailed tests of the numerical methods are presented in Sec.~\ref{sec:tests}. Our main findings for the three setups considered are presented in Sec.~\ref{sec:results}, in which we discuss and compare the properties of the numerically found optimal lawn shapes and the corresponding grasshopper success probabilities, as well as interpret these results in the context of the spherical harmonics representation. In Sec.~\ref{sec:conclusions}, we summarize our findings and connect them to our prior work and to other fields of research that exhibit similar pattern formation.

\section{Formal problem statement}
\label{sec:problemstatement}
The formal problem statement is similar to the planar case. The general grasshopper success probability functional for jump angle $\theta$ is given by
\begin{equation}
    p(\theta)=
    \frac{1}{4\pi^2\sin(\theta)}\int_{L_1}d^2r_1\int_{\overline{L}_2}d^2r_2\delta(\theta_{12}-\theta),\label{eq:continuous}
\end{equation}
where the $r_1$-integration goes over the first lawn $L_1\subset\mathbb{S}^2$, the $r_2$-integration goes over the complement of the second lawn $\overline{L}_2\subset\mathbb{S}^2$, and 
\begin{equation}
    \theta_{12}=2\arcsin(|\mathbf{r}_1-\mathbf{r}_2|/2)=\arccos(\mathbf{r}_1\cdot\mathbf{r}_2)
\end{equation}
is the spherical angle between the points $\mathbf{r}_1$ and $\mathbf{r}_2$ on the unit sphere $\mathbb{S}^2$. Both lawns cover exactly half of the surface area of the sphere. The goal is to determine optimal sets $L_1$ and $L_2$ for each value of $\theta\in[0,\pi]$ that maximize the probability $p(\theta)$.

In the general antipodal two-lawn setup, which we also call the independent antipodal setup, there are no other constraints on the sets $L_1$ and $L_2$. For antipodal complementary lawns, the constraint is $L_1=\overline{L}_2$. This is equivalent to demanding that the grasshopper must remain on the same lawn after jumping, as for the planar problem. This setup is therefore also referred to as the antipodal one-lawn setup. The final setup considered here also assumes $L_1=\overline{L}_2$, but without the antipodal condition. The only constraint is that both lawns must cover exactly half of the hemisphere. We call this setup the non-antipodal one-lawn setup or, equivalently, the non-antipodal complementary setup.

Another way of expressing the probability functional Eq.~\eqref{eq:continuous} is by introducing lawn density functions $\mu_k(\mathbf r)\in\{0,1\}$, defined for all points on the unit sphere, $\mathbf{r}\in\mathbb{S}^2$. Each lawn $L_k$ is defined as the set of points where $\mu_k(\mathbf{r})=1$ and its complement $\overline{L}_k$ is the set of points where $\mu_k(\mathbf{r})=0\Leftrightarrow\overline{\mu}_k(\mathbf{r})\equiv 1-\mu_k(\mathbf{r})=1$. Since the sphere has radius one, distances on the sphere and the corresponding spherical angles have the same magnitude, which leads to the normalization condition $\int_{\mathbb{S}^2}d^2r\mu_k(\mathbf r)=2\pi$. In antipodal setups we have the additional constraint $\mu_k(\mathbf r)+\mu_k(-\mathbf r)=1$ for all $\mathbf r\in\mathbb{S}^2$. In this notation, Eq.~\eqref{eq:continuous} takes the form
\begin{equation}
    p(\theta)=
    \frac{1}{4\pi^2\sin(\theta)}\int_{\mathbb{S}^2}d^2r_1\mu_1(\mathbf{r}_1)\int_{\mathbb{S}^2}d^2r_2\overline{\mu}_2(\mathbf{r}_2)\delta(\theta_{12}-\theta).\label{eq:continuousmu}
\end{equation}

In some special cases, this integral can be computed analytically. In particular, for complementary hemispherical lawns, the grasshopper success probability equals $1-\theta/\pi$ \cite{kent2014bloch}. For jump angle $\theta=\pi$ any antipodal lawn configuration has success probability equal to zero, since the grasshopper must jump from any point to its antipode. For jump angle $\theta=\pi/2$ any antipodal lawn configuration has success probability equal to $1/2$, since the grasshopper lands on a great circle exactly half of which is covered by the lawn by definition. For independent antipodal lawns, the symmetry of the integral implies that if $L_1$ and $L_2$ are an optimal lawn pair for jump angle $\theta$, then $L_1$ and $\overline{L}_2$ are an optimal lawn pair for jump angle $\bar{\theta}\equiv\pi-\theta$ with the same grasshopper success probability. 

Aside from these special cases and the analytical results from Refs.~\cite{kent2014bloch,chistikov2020globehopping}, which characterize discrete jump angles for which hemispherical lawns are optimal, the best presently available results on optimal lawn shapes are from numerical computations, as described in the following section.

To gain additional analytical insights, which will be discussed in Sec.~\ref{sec:harmonics}, it is convenient to recast \eqref{eq:continuousmu} using the normalized geodesic-circle averaging operator
\begin{equation}
(K_\theta f)(\mathbf r)
=\frac{1}{2\pi\sin\theta}\int_{\mathbb{S}^2}\delta\!\big(\theta(\mathbf r,\mathbf r')-\theta\big)\,f(\mathbf r')\,d^2r',
\end{equation}
for which $K_\theta 1=1$. Then
\begin{equation}
p(\theta)=\frac{1}{2\pi}\int_{\mathbb{S}^2}\mu_1(\mathbf r)\,(K_\theta \overline{\mu}_2)(\mathbf r)\,d^2r.
\label{eq:pK}
\end{equation}
We can now expand each lawn in orthonormal spherical harmonics \cite{AtkinsonHan2012},
\begin{eqnarray}
\mu_k(\mathbf r)&=&\sum_{\ell=0}^{\infty}\sum_{m=-\ell}^{\ell}\widehat\mu^{(k)}_{\ell m}\,Y_{\ell m}(\mathbf r),\\
\widehat\mu^{(k)}_{\ell m}&=&\int_{\mathbb{S}^2}\mu_k(\mathbf r)\,Y_{\ell m}^*(\mathbf r)\,d^2r.
\end{eqnarray}
The $\ell=0$ term is fixed by the total lawn area constraint: since $Y_{00}=1/\sqrt{4\pi}$ we obtain $\widehat\mu^{(k)}_{00}=\sqrt{\pi}$ for $k=1,2$. In the antipodal setups the relation $\mu_k(\mathbf r)+\mu_k(-\mathbf r)=1$ implies $\widehat\mu^{(k)}_{\ell m}=0$ for all even $\ell\ge2$. Additional constraints are imposed by the binary form of the lawn. Because $\mu_k(\mathbf{r})$ is real-valued we have $\widehat\mu^{(k)}_{\ell,-m}=(-1)^m\widehat\mu^{(k)*}_{\ell m}$, and because $\mu_k^2=\mu_k$ we have the spherical harmonics analog of Parseval's theorem,
\begin{equation}
\label{eq:parseval}
2\pi=\int_{\mathbb{S}^2}|\mu_k(\mathbf{r})|^2d^2r=\sum_{\ell,m}|\widehat\mu^{(k)}_{\ell m}|^2.
\end{equation}
Generally, the definition $\mu_k(\mathbf{r})\in\{0,1\}$ strongly constrains the allowed values of the coefficients $\widehat\mu^{(k)}_{\ell m}$.

The spherical harmonics representation yields a succinct form of the probability functional \eqref{eq:pK}. Because the integrand in the definition of the operator $K_\theta$ depends only on the relative angle between $\mathbf{r}$ and $\mathbf{r}'$, we can apply the Funk-Hecke formula \cite{AtkinsonHan2012} on $\mathbb{S}^2$, which says that for any function $f(\mathbf{r}\cdot\mathbf{r}')$ on the sphere,
\begin{equation}
    \int_{\mathbb{S}^2} f(\mathbf{r}\cdot\mathbf{r}') Y_{\ell m}(\mathbf{r}')d^2r' = 2\pi\int_{-1}^1 f(t)P_\ell(t)dt Y_{\ell m}(\mathbf{r}),
\end{equation}
where $P_\ell(t)$ is the Legendre polynomial of degree $\ell$. Hence $K_\theta$ is diagonal in the spherical harmonics,
\begin{equation}
K_\theta Y_{\ell m}=P_\ell(\cos\theta)\,Y_{\ell m},
\end{equation}
and inserting the expansions into \eqref{eq:pK} gives the spectral form
\begin{equation}
\label{eq:p-spectral-general}
p(\theta)=\frac{1}{2\pi}\sum_{\ell=0}^\infty\sum_{m=-\ell}^{\ell}
\widehat\mu^{(1)}_{\ell m}\,\widehat\mu^{(2)}_{\ell m}{}^{\!*} P_\ell(\cos\theta).
\end{equation}
In the one–lawn (complementary) case, $\overline\mu_2(\mathbf{r})=\mu_1(\mathbf{r})\equiv\mu(\mathbf{r})$, so \eqref{eq:p-spectral-general} reduces to
\begin{equation}
\label{eq:p-spectral-onelawn}
p(\theta)=\frac{1}{2\pi}\sum_{\ell=0}^\infty\sum_{m=-\ell}^{\ell}\big|\widehat\mu_{\ell m}\big|^2P_\ell(\cos\theta).
\end{equation}
Note that the $\ell=0$ term gives a contribution of $1/2$ to the probability in Eqs.~\eqref{eq:p-spectral-general} and \eqref{eq:p-spectral-onelawn}. In the case of antipodal lawns, the remaining sum is restricted to odd $\ell$. The spherical harmonics representation is somewhat analogous to the Fourier representation of the planar grasshopper problem discussed in Ref.~\cite{goulko2017grasshopper}.

\section{Numerical method}
\label{sec:method}
Our numerical setup is conceptually similar to the numerical setup for the planar version, which was described in detail in Ref.~\cite{goulko2017grasshopper}. In the following we outline the most important differences, which are due to the curvature of the spherical domain.

In contrast to the planar case, it is impossible to generate a uniform grid on the surface of the sphere that has a sufficiently large number of points, as regular sphere tilings are limited to the symmetries of the Platonic solids.
However, there are many methods to generate approximately uniform grids on the sphere. 
For this work, we have considered the following grid types:
\begin{itemize}
\item Symmetric spherical $t$-design grids \cite{Womersley2017grids, Womersley2018spheregrid}: sets of points on a sphere that accurately approximate integrals of polynomials up to degree $t$ over the spherical surface;
\item Hierarchical Equal Area isoLatitude Pixelization (HEALPix\footnote{http://healpix.sf.net}) grids \cite{Gorski2005Healpix, Zonca2019}: pixelization that subdivides the sphere such that each pixel covers the same surface area;
\item Goldberg polyhedron grids \cite{TEANBY20061442,VonLaven2015}: grids constructed from subdivided icosahedra (or other polyhedra), projected onto the sphere;
\item ``Coulomb" grids: custom generated grids constructed by placing charges on the surface of a sphere that repel one another.
\end{itemize}
We performed detailed tests of the accuracy of different discretization setups, which are presented in the next section. 

Based on the outcomes of these tests, we found that $t$-design grids have overall the best properties for our specific problem, and hence this type of grid was used for most numerical simulations. These grids provide an approximately uniform distribution of antipodal points on the sphere and exhibit the smallest discretization errors (at fixed resolution) among all tested grids. However, these grids must be generated through a separate numerical optimization process, which limits the maximal number of grid points available. We obtained the point sets used in this work from Ref.~\cite{Womersley2017grids}, where the maximal number of available points is $52{,}978$.

For small ($\theta\rightarrow0$) and large ($\theta\rightarrow\pi$) values of the jump, a larger number of grid sites is necessary to accurately resolve the jump. For such jump angles we employed HEALPix grids with up to $303{,}372$ sites. HEALPix grids consist of equal-area cells arranged on iso-latitude rings. They allow efficient hierarchical refinement and are thus computationally inexpensive to generate. As shown in the next section, they also achieve good overall accuracy, although slightly worse compared to the symmetric spherical $t$-design grids at comparable resolution. However, accuracy can be improved by increasing the resolution.

For a consistency check, we also compared with grids based on Goldberg polyhedra and with custom generated ``Coulomb" grids. The former possess a high degree of geometric symmetry, which can bias numerical results and lead to larger discretization errors, which do not improve with increasing resolution. While Coulomb grids are more accurate, they do not outperform HEALPix grids, and generating the custom point sets requires substantial additional numerical effort.

For a spherical grid with $N$ points, each point is associated with a grid cell area of $4\pi/N$ and we define the associated average lattice spacing via $h=\sqrt{4\pi/N}$.
The lawn density function $\mu_k(\mathbf{r})$ now becomes a discrete ``spin" variable $s_{ki}\in\{0,1\}$, defined at the grid points $i$.
The condition that half of the sphere is covered by the lawn then translates into $\sum_i s_{ki}=N/2=2\pi/h^2$. Integrals over the unit sphere become sums over the grid sites weighted by the average grid cell area $h^2$. The discrete setup also requires a smoothed approximation of the $\delta$-function in Eqs.~\eqref{eq:continuous} and \eqref{eq:continuousmu}, $\delta(r)\rightarrow\delta_h(r)=\phi(r/h)/h$. As in \cite{goulko2017grasshopper}, we choose 
\begin{equation}
    \phi \! \left(  \frac{\Delta x}{h} \right) = \left\{
\begin{array}{ll}
\frac{1}{4}\left(1+\cos(\frac{\pi\Delta x}{2h})\right) & {\rm if~} |\Delta x|/h\leq 2 \\
0 & {\rm if~} |\Delta x|/h\geq 2
\end{array}\right.,
\label{eq:discretedistancefn}
\end{equation}
following \cite{peskin2002deltafn}. Taken together, this results in the following discrete version of the grasshopper probability functional,
\begin{equation}
\discretep{s}{\theta} = \frac{4}{\sin(\theta) N^2 h}\sum_{i, j}  s_{1i} \overline{s}_{2j} \phi \left( \frac{ \theta_{ij} - \theta}{h} \right). \label{eq:spinham}
\end{equation}
In the independent lawn setup, we consider two separate spin configurations for the two lawns, while in the complementary one-lawn setup (both antipodal and non-antipodal) $\overline{s}_{2j}=s_{1j}$ and the first spin index can be dropped.
Otherwise, Eq.~\eqref{eq:spinham} has the same structure as the corresponding discrete probability in the planar case discussed in Ref.~\cite{goulko2017grasshopper}. The continuum limit is recovered by taking $h\rightarrow 0$ and $N\rightarrow\infty$ simultaneously.
\begin{figure}
    \centering
    \includegraphics[width=\columnwidth]{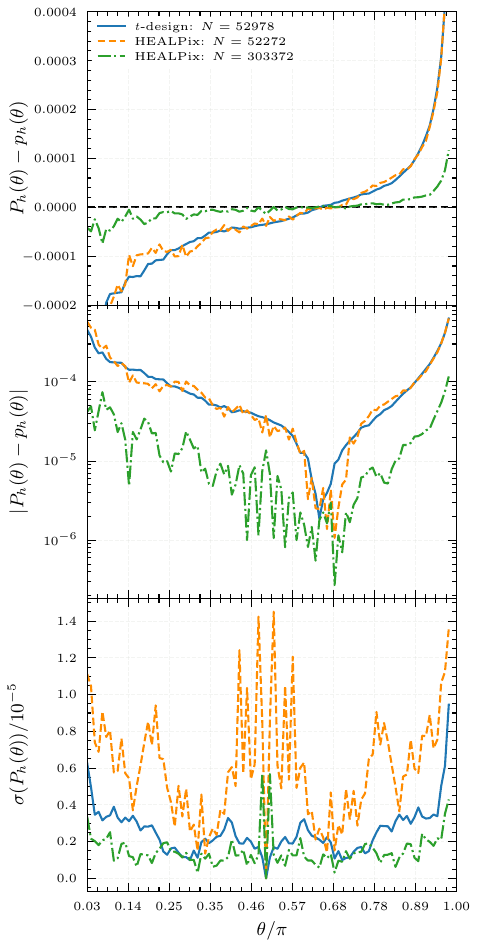}
    \caption{Study of discretization effects. The difference between the discrete probability ${P}_h(\theta)$ of a hemispherical lawn for different spherical grid types and sizes, and the exact continuous grasshopper probability $p_h(\theta) = 1 - \theta/\pi$ (top panel) and the corresponding absolute difference on a logarithmic scale (middle panel). 
    The discrete probabilities were obtained by averaging across a thousand random orientations of the hemispherical lawn relative to the grid.
    The width of the corresponding distribution (one standard deviation) is shown in the bottom panel. 
    The discretization errors are smaller than $0.1\%$ for most values of $\theta$.
    }
    \label{fig:hemisphere}
\end{figure}   

Formulated on a discrete grid the grasshopper problem corresponds to a conserved order parameter Ising model with Hamiltonian $H=-\discretep{s}{\theta}$. This Ising model has attractive fixed-range interactions, where the interaction range is very large in terms of the grid spacing. Pairs of spins only interact if the distance between them approximately corresponds to the grasshopper jump length $\theta$, which should be resolved by a large number of lattice spacings to ensure an accurate representation of the continuous model.
\begin{figure*}
    \centering
    \hfill
    \includegraphics[width=0.30\textwidth]{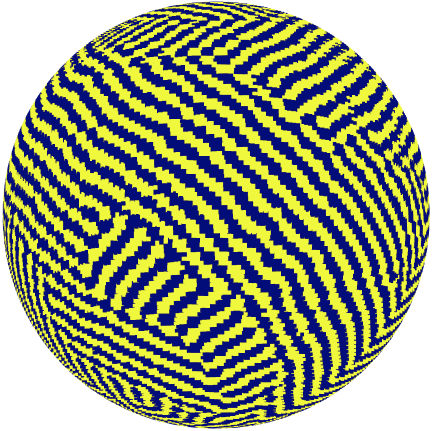}\hfill
    \includegraphics[width=0.30\textwidth]{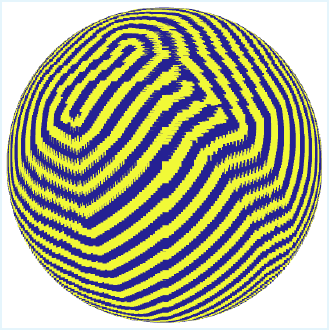}\hfill
    \includegraphics[width=0.30\textwidth]{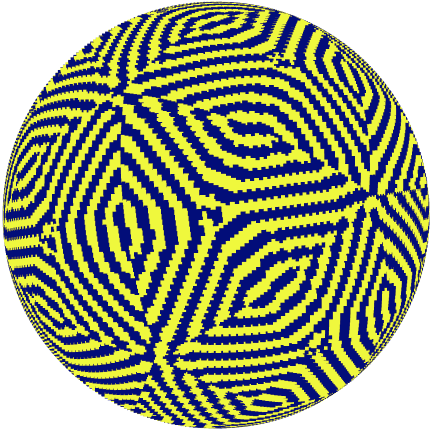}\hfill
    \caption{Optimal lawn configurations for $\theta = 0.98 \pi$ on the $t$-design grid (left), the HEALPix grid (middle), and the Goldberg grid (right). All grids have on the order of 40,000 points. It can be seen that the underlying symmetry of the Goldberg grid heavily influences the resulting optimal lawn shape.}
    \label{fig:bad_grid}
\end{figure*}

Global optimization techniques are required to numerically find discrete lawn configurations that correspond to optimal grasshopper lawns. An initial search for the global maximum of the discrete probability functional is performed using simulated annealing \cite{KiGeVe83}, following the same methodology as presented in Ref.~\cite{goulko2017grasshopper}. This search is then refined by annealing only the boundary of the system. Finally, a greedy algorithm is used to efficiently obtain the nearest maximum.

\section{Tests}
\label{sec:tests}

We performed extensive tests of the numerical setup. One important consistency check is reproducing the exact grasshopper success probability for a hemispherical lawn, $p_{h}(\theta) = 1 - \theta/ \pi$. Figure~\ref{fig:hemisphere} compares the continuous and discrete probabilities across a range of jumps for different grid types and grid sizes. Because the specific orientation of a given discrete hemispherical lawn with respect to the grid will slightly influence the corresponding probability, we generated a thousand discrete hemisphere lawns and averaged the results. The width of the corresponding distribution, shown in the bottom panel of Fig.~\ref{fig:hemisphere}, provides another consistency check: the discrete probability should change as little as possible when the orientation is changed.

The difference between the exact and the discrete hemisphere probabilities is below $0.1\%$ for most jumps. For all grid setups, discretization errors are largest for jump angles near $0$ and $\pi$, since in these limits the landing circle of the grasshopper is smallest and is thus resolved by only a few grid points. However, we can still get reasonably close to these limits with good accuracy. The discretization errors for HEALPix grids are comparable to the ones for $t$-design grids at similar resolution and are further improved by increasing the resolution. However, the probability varies more strongly with lawn orientation for HEALPix grids than for $t$-design grids with similar $N$.

Another concern is a potential alignment between lawn shapes and the features of the grid. Figure~\ref{fig:bad_grid} shows an example of the underlying symmetry of the grid influencing the lawn shape. This undesirable behavior is clearly pronounced with Goldberg grids, which have a high degree of symmetry. While we see some structures also in the $t$-design and HEALPix grids, the effect is much less pronounced and there is no regular pattern.

To evaluate how the irregularity of a spherical grid influences the lawn shape, we define the ``potential energy" of a grid point, $E_i$, to be the maximal amount (up to a prefactor) that a spin on that grid site can contribute to the overall discrete probability $\discretep{s}{\theta}$, i.e.\ the sum of of all possible interactions between the spin and grid points that it can interact with, if the latter were all occupied:
\begin{equation}
E_i = \sum _ { j } \phi \left( \frac{ \theta_{ij} - \theta}{h} \right).
\label{eq:potentialenergy}
\end{equation}
The potential energy of a grid point is only dependent on the grid itself, whereas the the specific contribution of a spin to the overall probability depends on the actual spin configuration. In a uniform infinite grid, every grid point would have the same potential energy. Hence, non-uniform grids that minimize the variance of the potential energies over all sites are favorable. The variance should also decrease with increasing number of grid points $N$. Another criterion is the regularity of the potential energy distribution, as we want to minimize the number of outliers with potential energies that deviate strongly from the mean.
Note that these metrics are customized for our specific numerical setup---standard measures of uniformity for spherical grids usually take into account local neighbors, while for the grasshopper setup interactions occur between grid points that are far away from each other and hence uniformity over larger scales is important. 

\begin{figure*}
    \centering
    \includegraphics[width=\textwidth]
    {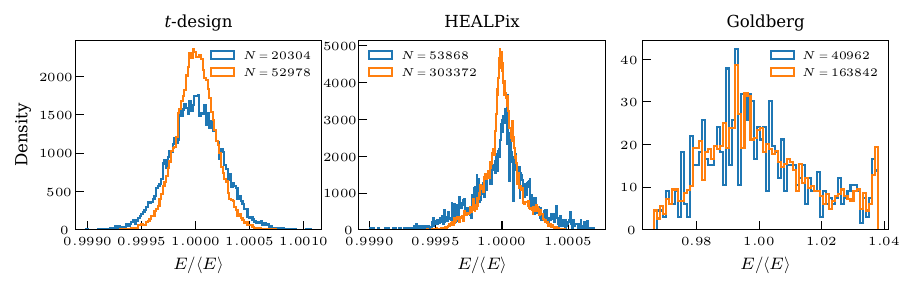}\hfill
    \caption{Centered histograms of potential energies \eqref{eq:potentialenergy} across the entire grid for $t$-design (left panel), HEALPix (middle panel), and Goldberg polyhedron grids (right panel) for the representative jump angle $\theta=0.30\pi$. The histograms for $t$-designs and HEALPix grids are more regular, with a width that decreases as the grid resolution $N$ is increased. The HEALPix grid histograms are more prominently peaked due to a higher concentration of points near the poles. The distributions for Goldberg polyhedron grids are much less regular with a larger overall width that does not appear to decrease with increasing $N$.}
    \label{fig:potenergies}
\end{figure*}
The corresponding numerical data for a representative value of $\theta$ are shown in Fig.~\ref{fig:potenergies}. The variance of the potential energy distribution for $t$-designs and HEALPix grids is significantly smaller than that for Goldberg polyhedron grids across all jump angles $\theta$. For $t$-designs and HEALPix grids, the variance decreases with increasing $N$, and their potential energy distributions are close to Gaussian. In contrast, the distributions for Goldberg polyhedra show no clear scaling with $N$ and are markedly less regular, featuring a long asymmetric tail.

Due to the above considerations, the numerical results presented in the following sections for jump angles $0.15\pi\leq\theta\leq0.80\pi$ were obtained using a $t$-design grid with $N=52{,}978$ sites, balancing accuracy and numerical efficiency. For jump angles outside of this interval a HEALPix grid with $N=303{,}372$ sites was used. We performed additional consistency checks between grid setups for select jump angles, to ensure that the qualitative properties of the optimal lawn shapes do not depend significantly on the type of grid.

\section{Results}
\label{sec:results}
In the following we present our numerical results for optimal grasshopper lawn shapes in the three setups: the antipodal complementary (one-lawn) setup, the antipodal independent (two-lawn) setup, and the non-antipodal complementary (one-lawn) setup. We analyze in detail the geometric properties of these lawn shapes. The corresponding grasshopper success probabilities are compared in Sec.~\ref{sec:probs} and we interpret our findings in terms of spherical harmonics in Sec.~\ref{sec:harmonics}.
\begin{figure*}   
    \includegraphics[width=\textwidth]{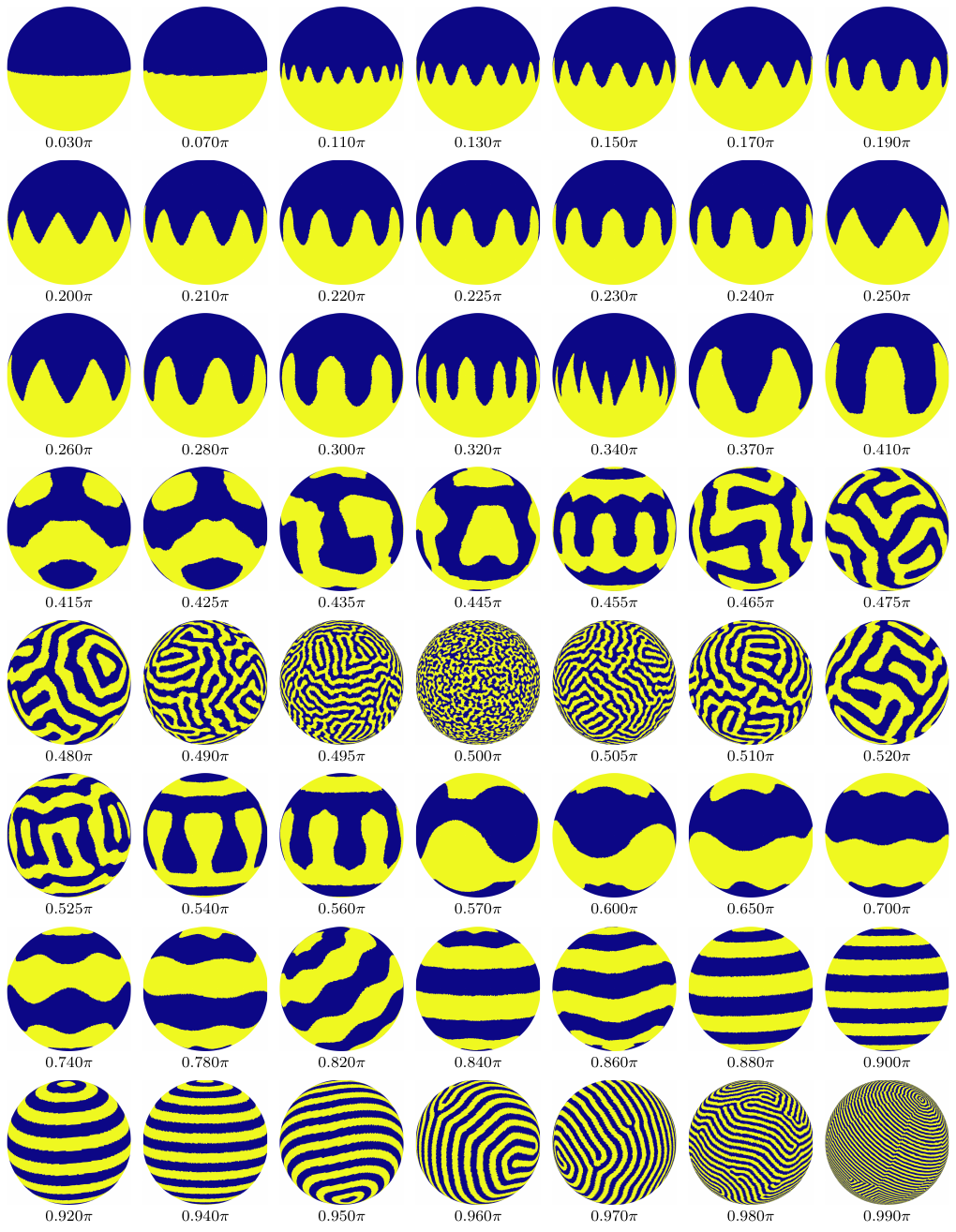}
    \caption{Optimal grasshopper spin configurations in the antipodal one-lawn setup for different values of the jump $\theta$.}
    \label{fig:antipodalshapes}
\end{figure*}

\subsection{Antipodal complementary (one-lawn) setup}
\label{sec:antipodalonelawn}
\begin{figure*}[!t]
    \includegraphics[width=\columnwidth]{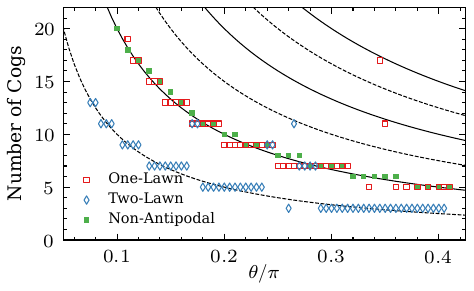}
    \includegraphics[width=\columnwidth]{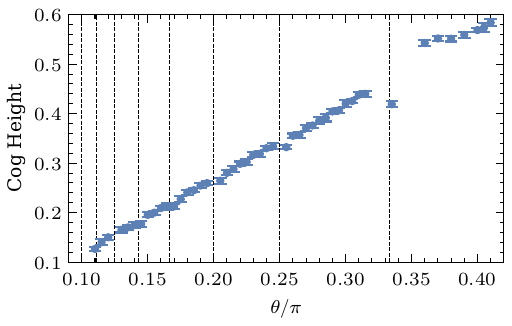}
    \caption{Left: Number of cogs in numerically found optimal lawn configurations in the cogwheel regime. The data for the antipodal one-lawn setup (red empty squares) is compared to the corresponding data for the non-antipodal one-lawn setup (green filled squares) and for the antipodal two-lawn setup (blue empty diamonds).
    In antipodal setups the number of cogs must be odd, while for non-antipodal lawns the number of cogs can be any integer. 
    Solid black lines correspond to multiples of $2\pi/\theta$ -- these modes are accessible in all setups. Dashed black lines show additional modes that are accessible only in the two-lawn setup, corresponding to odd multiples of $\pi/\theta$.
    Most numerically found optimal configurations belong to the lowest allowed mode. Right: Height of cogs in numerically found optimal lawn configurations in the antipodal one-lawn setup. Vertical dashed lines denote jump angles of the form $\theta=\theta_q=\pi/q$ for integer $q$.}
    \label{fig:cognumber}
\end{figure*}
\begin{figure*}
    \includegraphics[width=\columnwidth]{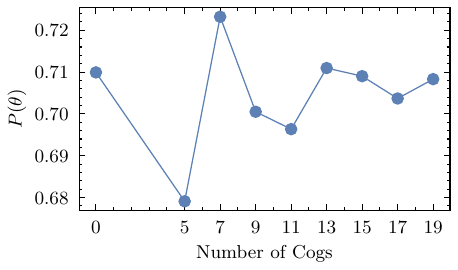}
    \includegraphics[width=\columnwidth]{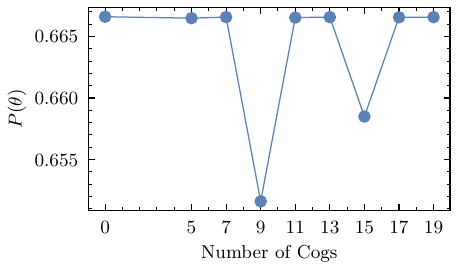}
    \caption{Discrete grasshopper success probability in the antipodal one-lawn setup for a generic jump angle $\theta=0.290\pi$ (left) and for $\theta=\theta_3=\pi/3$ (right) as function of the number of cogs. For this, the number of cogs was fixed by enforcing the corresponding symmetry and the specific shape was subsequently numerically optimized. For $\theta=0.290\pi$ the first and second modes (corresponding to 7 and 13 cogs, respectively) attain the two highest probabilities. For $\theta=\pi/3$ there are multiple near-degenerate probability maxima, including the hemispherical lawn. For this jump the first mode has either 5 or 7 cogs, the second mode either 11 or 13 cogs, the third mode either 17 or 19 cogs, and so on.}
    \label{fig:modesprobability}
\end{figure*}
We begin by discussing the antipodal one-lawn setup, in which the grasshopper aims to remain on the same lawn after the jump.
Figure~\ref{fig:antipodalshapes} gives an overview over the optimal lawn shapes across all values of the grasshopper jump angle $\theta$. We identify several regimes. For $\theta\lesssim 0.41\pi$ (first three rows in Fig.~\ref{fig:antipodalshapes}) the numerically found optimal shapes look like cogwheels, paralleling the planar case. At larger jumps the lawn shapes become irregular. Some exhibit bands with round or heart-shaped islands, or a combination of rings and cog-like features. Eventually, in the interval $0.47\pi\lesssim\theta\lesssim0.53\pi$ centered around $\pi/2$, the lawns resemble labyrinths, 
with finer and finer structures as $\theta$ approaches $\pi/2$.
As jump size is further increased, the configurations display a more regular combination of cogs and stripes, and for $\theta\gtrsim0.57\pi$ exclusively striped patterns. The boundaries of the stripes in this regime show cog-like modulations, which have higher frequency and smaller amplitude as $\theta$ is increased. The number of stripes also increases with $\theta$. Near $\theta=\pi$ the numerically generated stripe configurations start exhibiting defects. In the following we will discuss the features of these regimes in more detail.

\subsubsection{Cogwheel regime}
\begin{figure*}
    \centering
    \includegraphics[width=\textwidth]{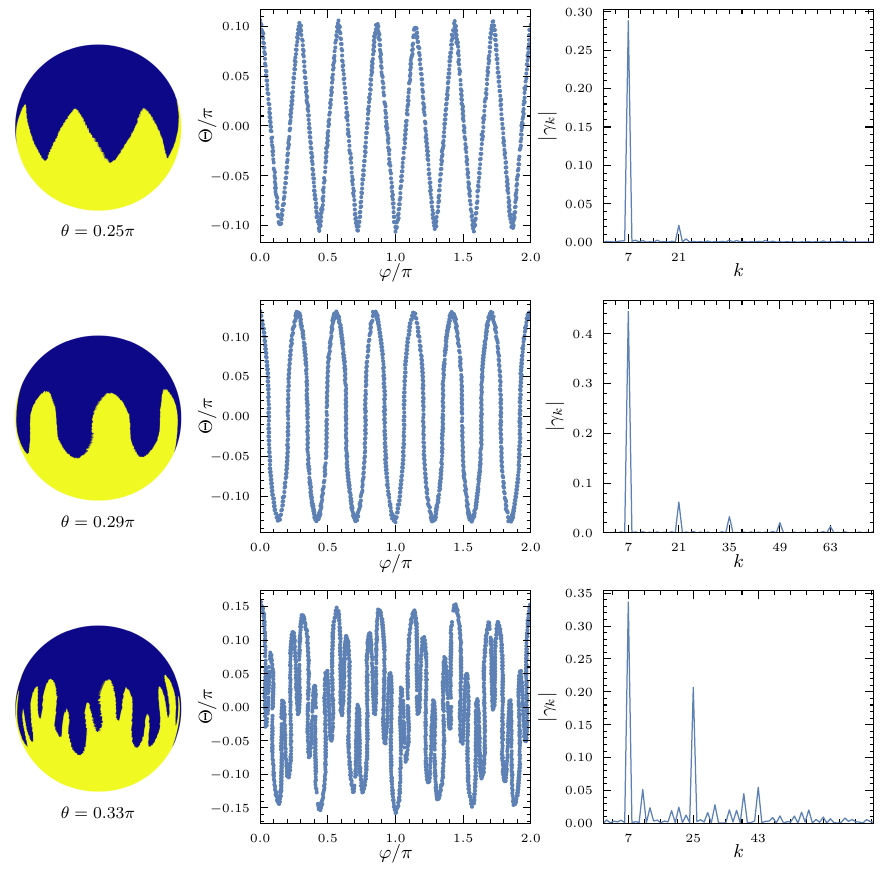}
    \caption{
    Optimal configurations in the antipodal one-lawn setup for $\theta$ values of $0.25\pi$, $0.29\pi$, and $0.33\pi$ (from top to bottom). Left panels: snapshots of the configurations. Middle panels: corresponding lawn boundaries in terms of the polar angle $\Theta$ as a function of the azimuthal angle $\varphi$. Right panels: Fourier amplitude spectra of the lawn boundaries.
    }
    \label{fig:fourier}
\end{figure*}

We first discuss the cogwheel regime. Due to the antipodal condition, the number of cogs must be odd. Specifically, we expect the distance between cogs to be close to the jump angle, which implies that the number of cogs is the odd integer closest to $2\pi/\theta$. This is somewhat analogous to the planar case, where the number of cogs is determined by the dihedral symmetry of a polygon inscribed in the unit area disk with edge length that is nearest to the grasshopper jump length. However, higher modes, i.e.\ configurations with a larger number of cogs, have also been observed numerically. These constitute at least local probability maxima. The number of cogs for mode $m$ is then given by the odd integer closest to $2m\pi/\theta$, so that $m$ can be interpreted as the winding number around the equator. The number of cogs for numerically found optimal lawn shapes is plotted in Fig.~\ref{fig:cognumber}, together with the corresponding cog numbers for the other two setups, which will be discussed in more detail in the following sections. Almost all numerically generated shapes in the antipodal one-lawn setup correspond to the first mode.
Figure~\ref{fig:cognumber} also shows the corresponding cog heights in the antipodal one-lawn regime. The cog heights are measured by extracting the boundary of the lawn as shown in Fig.~\ref{fig:fourier} and then computing the amplitude of the oscillations (the average distance of the peaks or troughs from the equator, with the error corresponding to one standard deviation).
The cog height increases approximately linearly with increasing $\theta$, but with some modulations. In particular, near $\theta=\theta_q\equiv\pi/q$, where $q$ is an integer, the cog heights tend to be somewhat lower. We discuss the variations in cog shapes as function of jump angle below. Note that around $\theta=\theta_3=\pi/3$ the data is sparse because most numerically generated configurations correspond to a mixed mode, rather than the first mode.

\begin{figure}
        \includegraphics[width=\columnwidth]{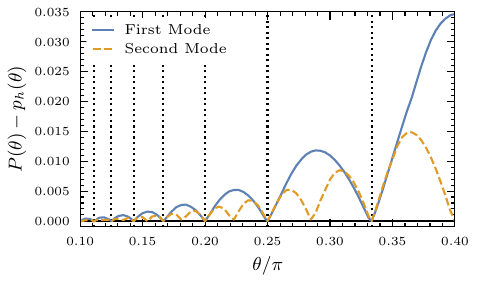}
    \caption{Probability difference between the cogwheel lawn (first mode: blue solid line, second mode: orange dashed line) and the hemisphere lawn as a function of $\theta$ in the antipodal one-lawn setup. Vertical dotted lines denote jump angles of the form $\theta=\theta_q=\pi/q$. The probability difference approaches zero at these jump angles. For small values of $\theta$ no data could be obtained for higher modes.}
    \label{fig:first-second-mode}
\end{figure}
An important special case is the presence of the $m=0$ mode for certain values of the jump angle.
It was shown in Ref.~\cite{chistikov2020globehopping} that the hemispherical lawn is not optimal, except for jump angles of the form $\theta=\theta_q\equiv\pi/q$, where $q$ is an integer. For the latter, it was previously known \cite{kent2014bloch} that the optimal probability is $1-1/q$ and this probability is attained by hemispherical lawns. Our numerical results correctly reproduce these analytical results. However, numerical results show that in addition to hemispherical lawns, there are also cogwheel-shaped lawns that approximate this optimal probability, indicating a degenerate or near-degenerate probability maximum.

\begin{figure*}
    \centering
    \includegraphics[width=\textwidth]{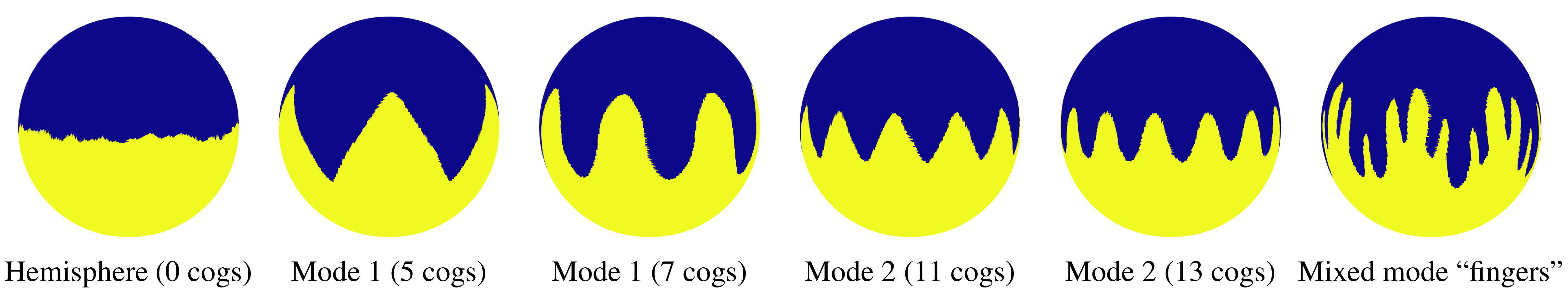}
    \caption{Lawn shapes for $\theta = \theta_3 = \pi/3$. From left to right: hemisphere lawn (0 cogs), mode 1 lawns (5 and 7 cogs), mode 2 lawns (11 and 13 cogs), and a mixed-mode configuration with a finger-like structure. All these shapes have similar grasshopper success probability.}
    \label{fig:modesshapes}
\end{figure*}
The grasshopper success probability as a function of the number of cogs for two representative values of the jump angle is shown in Fig.~\ref{fig:modesprobability}. For this analysis the number of cogs was fixed by enforcing the corresponding symmetry and the specific lawn shape was subsequently numerically optimized under this constraint. For generic jump angles the first mode has the highest probability, the second mode has the second-highest probability, etc. However for angles of the form $\theta=\theta_q$ there are multiple near-degenerate probability maxima, including the hemispherical lawn (corresponding to zero cogs). Note that in this case, for mode $m$ the optimal number of cogs approximates $2m\pi/\theta_q=2mq$, which is an even integer. Since the actual number of cogs in the antipodal setup must be odd, there are two good approximations on either side of this integer, $2mq\pm1$. 
Hence, each mode corresponds to two cog numbers with similar probability. A direct comparison between the probabilities of the first and second cogwheel mode and the hemisphere lawn probability as a function of jump angle is presented in Fig.~\ref{fig:first-second-mode}. Figure~\ref{fig:modesshapes} shows the lawn shapes for $\theta=\theta_3=\pi/3$ for the various modes. Note that a mixing of modes is also possible and represents at least a local probability maximum as the probability is approximately the same as that for the hemispherical lawn. The lawn boundary has a finger-like structure in this case.

The shapes of the cogs in the optimal configurations evolve with $\theta$. 
The second row of Fig.~\ref{fig:antipodalshapes} exemplifies this. It shows optimal first-mode configurations with $\theta$-values between $0.200\pi$ on the left (corresponding to $2\pi/\theta = 10$) and $0.250\pi$ on the right (corresponding to $2\pi/\theta = 8$). For the shape in the center, $\theta=0.225\pi$, i.e.\ $2\pi/\theta\approx9$.
If $2\pi/\theta$ is close to an even integer the cogs have a triangular shape, while for values $2\pi/\theta$ close to odd integers the cogs are rounder. 
It is interesting to compare these shape variations with the corresponding configuration of the planar grasshopper problem \cite{goulko2017grasshopper}. In the planar case, the respective widths of the teeth and gaps vary depending on how well the jump distance approximates the value given by the corresponding dihedral symmetry group. 
In the antipodal spherical case, the teeth and gaps need to have the same shape, but this shape itself varies depending on how close $2\pi/\theta$ is to an odd integer.

For a more detailed analysis of the cog shapes in several representative cases, Fig.~\ref{fig:fourier} shows different projections of the lawn configurations together with the Fourier amplitude spectrum of the lawn boundaries. For triangular cogs, the Fourier coefficients decline much more quickly than for the rounder cogs (the coefficients in the Fourier series for a regular triangle wave scale like $1/n^2$, while for a square wave they scale like $1/n$). For the mixed mode there are several prominent peaks, in addition to multiple smaller peaks.

Jumps of size $\theta_q$ are of particular interest in the one-lawn setup. As discussed above, for these jump sizes the hemispherical lawn is optimal yet our numerical results favor cogged shapes. The antipodal condition prohibits the ``natural'' choice of $2q$ cogs, which corresponds to a jump of $\theta_q$ between adjacent cogs, but our results give configurations with $2q-1$ or $2q+1$ cogs that seem to have near optimal success probabilities. A remarkable feature of these lawns is that the cogs appear to have shapes very close to triangular. 

We can take advantage of the symmetry and simple geometry of such lawns for a more exact analysis of the associated probabilities as a function of the height of the cogs. For any point $p$ on the sphere we have trigonometric equations for the intersections of the boundaries of the cogs with the circle on the sphere representing a jump of $\theta_q$ from $p$. From these we can compute the success probability for that point with arbitrary precision, and by numerical integration over the lawn we can get the success probability for jumps on the lawn with high accuracy. Details of the calculation are given in Appendix~\ref{sec:appendix}.

\begin{figure}
\includegraphics[width=\columnwidth]{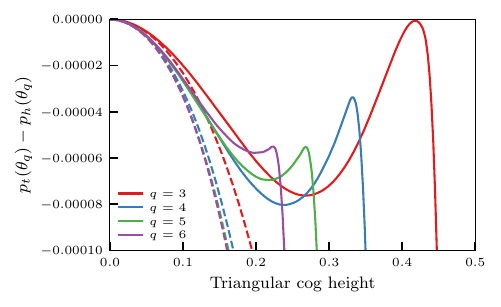}
\caption{Success deficit (defined as the success probability of the triangular cogged lawn minus the success probability of the hemispherical lawn) as function of cog height for different jump angles $\theta_q$. Solid curves correspond to lawns with $2q-1$ cogs and dashed curves to lawns with $2q+1$ cogs. }
\label{fig:triangularresults}
\end{figure}

Our results for $q=3,\ldots,6$ with $2q\pm1$ cogs are shown in Fig.~\ref{fig:triangularresults}. What we call the ``success deficit" is the success probability for the cogged lawn minus the optimal probability of $1-1/q$ attained by the hemisphere. 
These show that the success probabilities for triangular cogs are slightly worse than for the hemisphere, but that the differences are extremely small until the cog height becomes large. While for configurations with $k=2q+1$ cogs the size of the success deficit grows monotonically with increasing cog height, for $2q-1$ cogs there is a clear local maximum success probability at positive cog height for each $q$ plotted. In particular, for $q=3$ the size of the deficit at the local maximum is less than $10^{-6}$, so the hemisphere and the triangular cogged lawn are almost equivalent in terms of success probability. The cog heights at which these local maxima occur are very close to the cog heights of the numerically found optimal discrete configurations that are shown in Fig.~\ref{fig:cognumber}.

\subsubsection{Labyrinth regime}
We next discuss the labyrinth regime that occurs around $\theta=\pi/2$. As mentioned previously, at this jump angle all lawn configurations have the same grasshopper success probability, $1/2$. This angle is a special case of $\theta_q$ with $q=2$ (and hence $1-1/q=1/2$). As we approach this critical angle from either direction, the optimal lawn shapes display labyrinth-like formations. The fineness and complexity of the labyrinths increase closer to the critical angle. Simulating exactly at $\theta=\pi/2$ results in completely random lawns, as expected, despite discretization effects. The fifth row of Fig.~\ref{fig:antipodalshapes} shows the corresponding labyrinthine shapes around and at $\theta=\pi/2$.

\subsubsection{Stripes regime}
\begin{figure}
    \centering
    \includegraphics[width=\columnwidth]{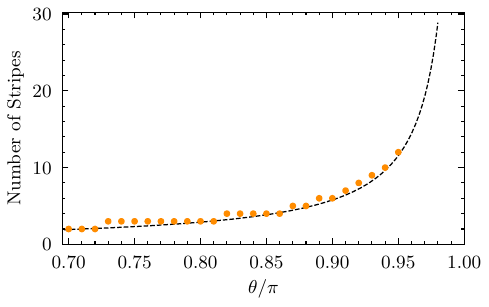}\hfill
    \caption{The number of stripes in numerically found optimal configurations in the antipodal one-lawn setup as function of the jump angle $\theta$. The dashed line is the analytical prediction from Eq.~\eqref{eq:stripenumber}.}
    \label{fig:stripes}
\end{figure}

For larger jump angles, $\theta\gtrsim0.57\pi$, the optimal shapes exhibit stripes that wind around the sphere. These stripes, or rings, are parallel until the jump angle becomes close to $\pi$. The number of stripes increases with increasing $\theta$ (and thus the width of the stripes decreases). The boundaries of the stripes also exhibit cog-like perturbations. The amplitude of these perturbations decreases with increasing $\theta$, while their number increases, making them less pronounced as $\theta$ approaches $\pi$. The number of stripes is plotted in Fig.~\ref{fig:stripes}.

As we show below, the number of stripes $n_s$ is well described by the analytical expression 
\begin{equation}
n_s = \frac{\pi}{\sqrt{3} \left(\pi - \theta\right)} ,
\label{eq:stripenumber}
\end{equation}
which can be derived by considering a planar approximation to the problem. 

\begin{figure}
    \centering
    \includegraphics[width=\columnwidth]{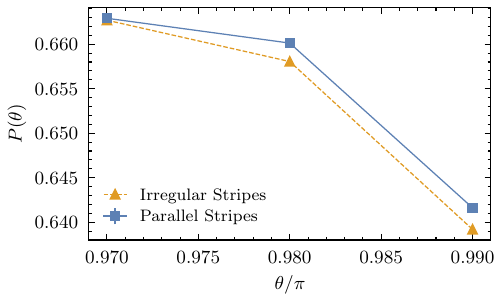}
    \caption{
    Probabilities for irregular striped configurations produced by numerical optimization (orange triangles) and probabilities for corresponding explicitly constructed parallel-stripe configurations (blue squares) for several values of jump angle $\theta$ near $\pi$ on a HEALPix lattice with 303,372 sites.}
    \label{fig:stripe_comparison}
\end{figure}

For very large jumps, $\theta\gtrsim 0.95\pi$, simulated annealing produces shapes where the stripes are ``broken", developing irregularities, such as splitting and bending. These irregular striped lawns are qualitatively different from the labyrinth-shaped lawns that occur near $\theta=\pi/2$. To assess whether these defects are a numerical artifact or a property of the optimal solutions, we compare the success probabilities of the irregular configurations produced by numerical optimization against explicitly constructed regular stripe patterns with matching stripe widths. The result of this comparison is shown in Fig.~\ref{fig:stripe_comparison}. The regular stripe configurations tend to have higher success probability than the irregular ones, but the difference is very small. 
It is plausible that irregular configurations are generated more frequently during the simulated annealing process, since they have almost the same success probability as their regular counterparts, while their multiplicity is substantially larger.

In the following, we attempt to explain these configurations and analyze the width of stripes as a function of the jump angle. An exact characterization of the optimal configurations is beyond our ambitions so our aim is just to present a reasoned explanation of the observed numerical data.  

The antipodal condition gives us immediately that a jump of size $\theta$ from the lawn to the lawn corresponds exactly to a jump of size $\bar{\theta}=\pi - \theta$ from the lawn to non-lawn. Therefore the case that we are considering can be regarded as making tiny jumps from lawn to non-lawn. Furthermore, for the analysis of optimal patterns for tiny jumps it is reasonable to approximate small patches of the spherical surface by a plane. To try to understand the observed local striping and predict the width of these stripes we undertake an exact analysis of taking small jumps on a plane composed of parallel blue and yellow stripes of width $w$. We calculate the probability of a successful jump from a random blue point in a random direction and landing on a yellow point. 

\begin{figure}
\begin{center}
\includegraphics[width=\columnwidth]{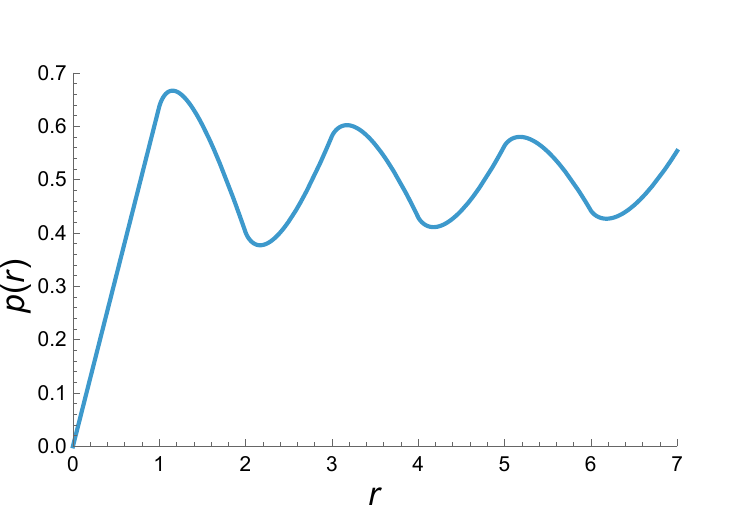}
\caption{Grasshopper success probability in the plane for stripes of unit width as a function of the jump size $r$.}
\label{fig:stripes-success1}
\end{center}
\end{figure}

For calculating the success probability on the plane, we need only consider the ratio $r$ between the jump size, $\bar{\theta}$, and the stripe width, so it is notationally convenient to take the stripe width to be $1$ and consider jumps of size $r$. It is clear that when $r$ is large, corresponding to the grasshopper's landing circle being of large radius superimposed on a pattern of thin parallel stripes, the success probability approaches $1/2$. A simple numerical calculation shows that the value of $r$ giving optimal success probability is less than~$2$, see Fig.~\ref{fig:stripes-success1}. 

\begin{figure}
\begin{center}
\includegraphics[width=\columnwidth]{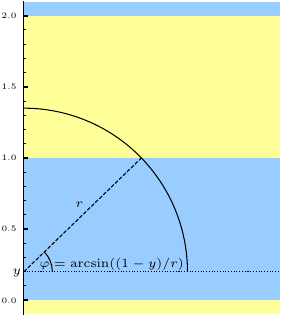}
\caption{Sketch of the stripes in the planar approximation with jump size $r\approx 1.15$. Without loss of generality, the grasshopper starting point is chosen to be $(0,y)$ on the blue stripe and the direction of the jump is given by the angle $\varphi$. Success is represented by a jump from the lawn (blue) to non-lawn (yellow). The critical angle $\varphi$ for a successful jump is $\arcsin\left(\frac{1-y}{r}\right)$.} 
\label{fig:stripes-sketch}
\end{center}
\end{figure}

\begin{figure}
\begin{center}
\includegraphics[width=\columnwidth]{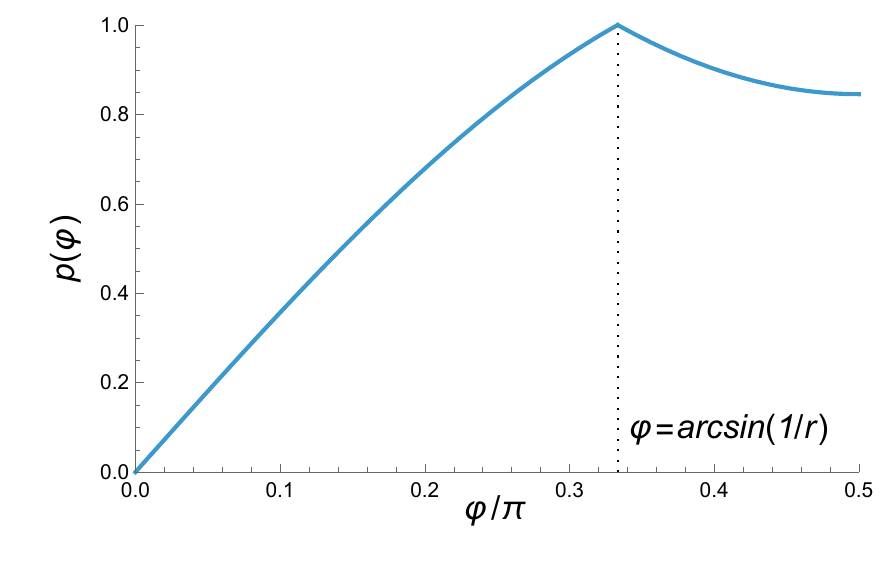}
\caption{Grasshopper success probability for the stripes regime for $r\approx1.15$ and averaged over all possible starting offsets $0\leq y < 1$ as a function of the jump orientation $\varphi$ in the planar approximation. The success probability is optimal for $\varphi=\arcsin(1/r)$.}
\label{fig:critical-stripe-angle}
\end{center}
\end{figure}

Once we know that the optimal value of $r$ is less than~$2$, the analysis is straightforward. Without loss of generality, we assume the stripes to be horizontal, i.e.\ parallel to the $x$-axis, and define the grasshopper starting point to be $(0,y)$ with $0\leq y < 1$. We introduce the angle $\varphi$ starting from the positive $x$-axis and going anticlockwise, which characterizes the direction of the grasshopper jump relative to the orientation of the stripes, as sketched in Fig.~\ref{fig:stripes-sketch}. By symmetry we can limit the range of $\varphi$ from~$0$ (along the $x$-axis) to $\pi/2$ (along the $y$-axis). With $\varphi=0$ the success probability is zero since the grasshopper lands on the same color as its take-off point. If $r\ge 1$ and $\varphi=\arcsin(1/r)$ then the success probability is $1$ for any $y$, since the landing spot is exactly distance $1$ above the take-off spot and so is of the opposite color. Fig.~\ref{fig:critical-stripe-angle} shows, with $r\approx 1.15$ for example\footnote{To be precise, the jump size chosen for this example is $r=2/\sqrt{3}\approx1.15$, which, as we will later show, is the optimal value.}, the probability of the grasshopper landing on an opposite colored stripe for $y$ chosen uniformly at random over $[0,1)$, as a function of $\varphi$.  

\begin{figure}
\begin{center}
\includegraphics[width=\columnwidth]{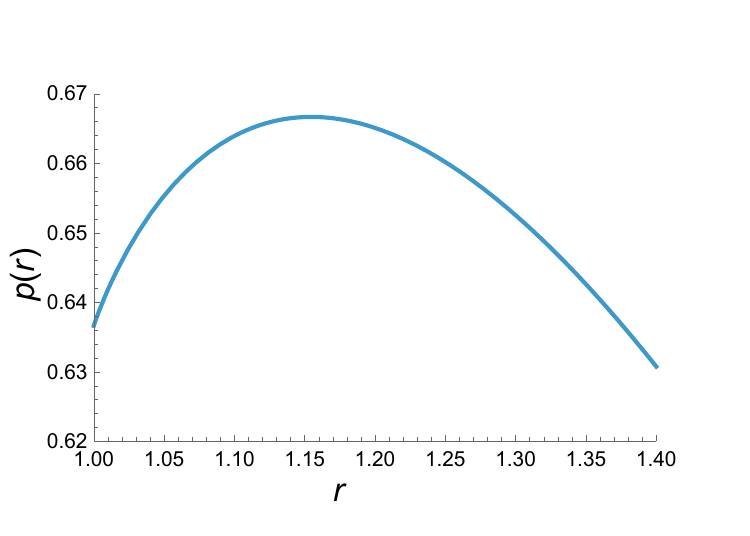}
\caption{Success probability in the planar approximation as a function of jump size $r$ near optimality. The function has a clear maximum.}
\label{fig:stripes-success2}
\end{center}
\end{figure}

We can now compute the overall success probability by integrating over $\varphi$. The grasshopper is successful if the $y$-coordinate of its landing point falls between $1$ and $2$, which corresponds to $1-r\sin\varphi\leq y<2-r\sin\varphi$. This has probability
$h(r,\varphi)= r \sin \varphi$ if $r \sin\varphi\le 1$ and $2-r \sin \varphi$ if $r \sin \varphi > 1$, which is the function $h(r,\varphi)$ shown in Fig.~\ref{fig:critical-stripe-angle}. Each of these expressions can be integrated explicitly, 
so if $\alpha_r = \arcsin\frac1{r}$ then
\begin{eqnarray}
\int_{0}^{\pi/2}h(r,\varphi)\ d\varphi &=& \int_{0}^{\varphi_r}r \sin \varphi\ d\varphi + \int_{\varphi_r}^{\pi/2}(2-r \sin \varphi)\ d\varphi\nonumber\\
&=& r-\sqrt{r^2-1} + \pi -\sqrt{r^2-1} - 2\varphi_r .
\end{eqnarray}
The integral is over the range $[0,\pi/2]$, so the total success probability as a function of $r$ is given by $u(r)= \frac2{\pi}\int_{0}^{\pi/2} h(r,\varphi) d\varphi $. 
This function $u(r)$ is shown in Fig.~\ref{fig:stripes-success2} and has a clear maximum over this range.
Now, $\frac{d}{dr} u(r) = 1-2\sqrt{1-1/r^2}$ so the maximum value is at $r=2/\sqrt{3}$ and $u(2/\sqrt{3}) = 2/3$.

It seems that as $\theta$ approaches $\pi$ we can define lawn patterns consisting of local patches of parallel stripes such that the success probability approaches $2/3$. The patches should be small enough so that they 
are close enough to planarity for their striping, and large enough so that the confused areas at patch boundaries represent a negligible proportion of the total area. 
The stripe width is approximately $\frac{\sqrt{3}}{2}(\pi - \theta)$ yielding the approximation for the number of stripes of each color, $n_s=\pi/2w$, quoted in Eq.~\eqref{eq:stripenumber}.
Interestingly, while the limit of the grasshopper success probability as $\theta\rightarrow\pi$ seems to be $2/3$, the value of the probability at $\theta=\pi$ exactly is zero due to the antipodal condition. 
Thus, there is apparently a discontinuity as $\theta\rightarrow\pi$. 

We have no proof that values greater than $2/3$ 
are unachievable though this seems likely since our numerical calculations do not improve on this value. In view of the very simple nature of our $2/3$ lower bound, it would be of interest to find a simple proof of this which avoids detailed integrals. 

This is reminiscent of Buffon's needle problem (see~\cite{buffon1733, buffon1777} and, e.g.,~\cite[Sec.~1.1]{mathai99}): If a plane is divided into equally spaced parallel stripes, what is the probability that a needle dropped randomly on the plane will cross the boundary between two stripes?

\subsection{Antipodal independent (two-lawn) setup}
\label{sec:antipodaltwolawn}
\begin{figure*}[!htbp]
    \includegraphics[width=\textwidth]{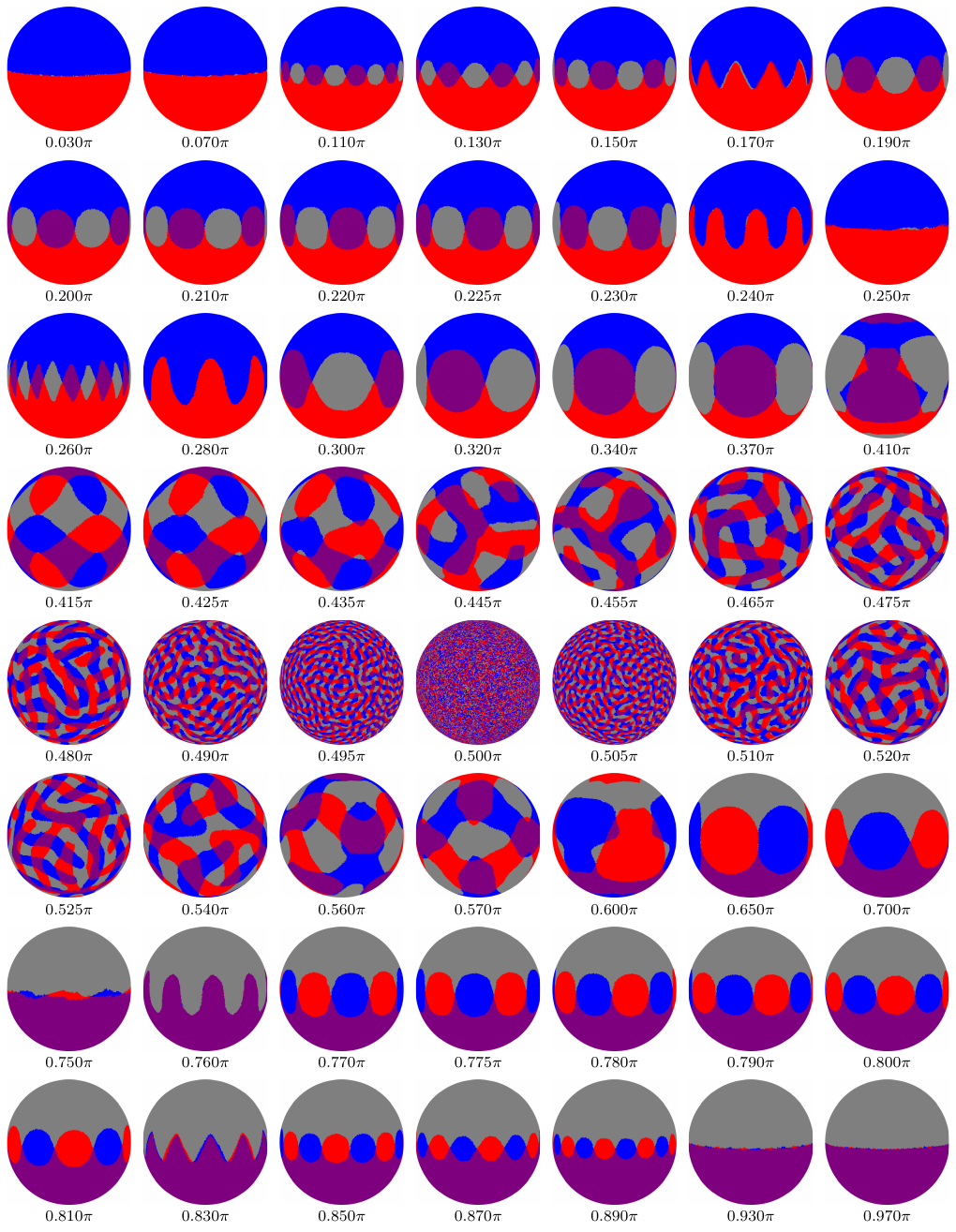}
    \caption{Optimal grasshopper spin configurations in the antipodal two-lawn setup for different values of the jump $\theta$. The two lawns are represented by the red and blue colors; gray areas are not covered by either lawn and purple areas are covered by both lawns.}
    \label{fig:two-coloring-shapes}
\end{figure*}

In the more general setup, we allow for two independent antipodal lawns, where the grasshopper starts on the first lawn and then jumps such that success corresponds to the grasshopper landing \textit{outside of} the second lawn. 
Numerically found optimal grasshopper shapes in this setup for different jumps are shown in Fig.~\ref{fig:two-coloring-shapes}.

\begin{figure*}
    \centering
    \includegraphics[width=\textwidth]{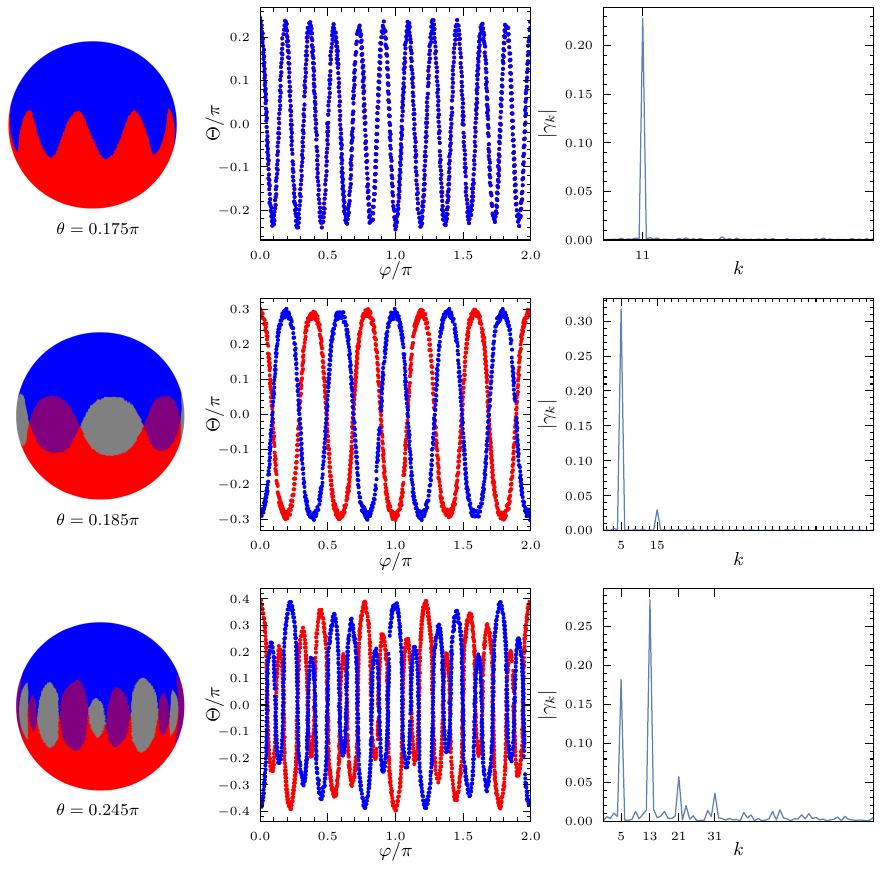}
    \caption{Optimal configurations in the two-lawn setup for $\theta = 0.175\pi$, $0.185\pi$, and $0.245\pi$ (top to bottom).
Left panels: snapshots of the optimal configurations on the sphere.
Middle panels: corresponding lawn boundaries expressed as the polar angle $\Theta(\varphi)$ as a function of the azimuthal angle $\varphi$.
Right panels: Fourier amplitude spectra $|\gamma_k|$ of the boundary shapes.
For $\theta = 0.175\pi$, the optimal configuration coincides with the one-lawn solution and exhibits an in-phase boundary.
At $\theta = 0.185\pi$, the two lawns adopt an out-of-phase configuration, visible as the two lawns overlapping.
This out-of-phase structure allows the two-lawn system to occupy lower modes that are inaccessible in the one-lawn setup, reducing the optimal number of cogs.}
    \label{fig:independent-modes}
\end{figure*}

For jumps below $\pi/2$ the qualitative regimes are the same as in the one-lawn setup: cogwheel-like structures (or hemispheres for certain jump angles, as discussed below), followed by an intermediate regime and eventually labyrinths near $\theta=\pi/2$. However, the number of cogs now has a different pattern, as can be seen in Fig.~\ref{fig:cognumber}. The two independent lawns in this setup have the same numbers and shapes of cogs, but for the first mode, which corresponds to an optimal solution, the number of cogs is the odd integer nearest to $\pi/\theta$ (as opposed to $2\pi/\theta$ for the one-lawn setup) and the lawns are offset from each other by half the distance between the cogs. For the general mode $m$ in the two-lawn setup the number of cogs is approximated by $m\pi/\theta$. This implies the existence of additional modes compared to the one-lawn case. For odd values of $m$ the lawns are offset and for even values of $m$ the lawns are exactly complementary, see Fig.~\ref{fig:independent-modes} for a representative example. The latter are the same configurations that emerge in the one-lawn setup. The additional freedom to offset modes allows for higher grasshopper success probabilities in the two-lawn setup. In contrast to the one-lawn case, for special jump angles $\theta_q=\pi/q$ the number of cogs in the first mode is now given by $q$ (rather than $2q$) and this number can be either even or odd. For even $q$ rounding to the nearest odd integer is required. In this case the hemisphere is an optimal (albeit near-degenerate) solution to the two-lawn grasshopper problem, like in the one-lawn case. For odd $q$ on the other hand, there exist cogwheel solutions that have higher success probabilities than the hemisphere. These findings confirm prior analytical results that for two independent antipodal lawns the hemisphere is optimal for $\theta=\theta_q$ and $q$ even \cite{kent2014bloch} but not for any other jump angle \cite{chistikov2020globehopping}.
\begin{figure*}[!htbp]
    \includegraphics[width=\textwidth]{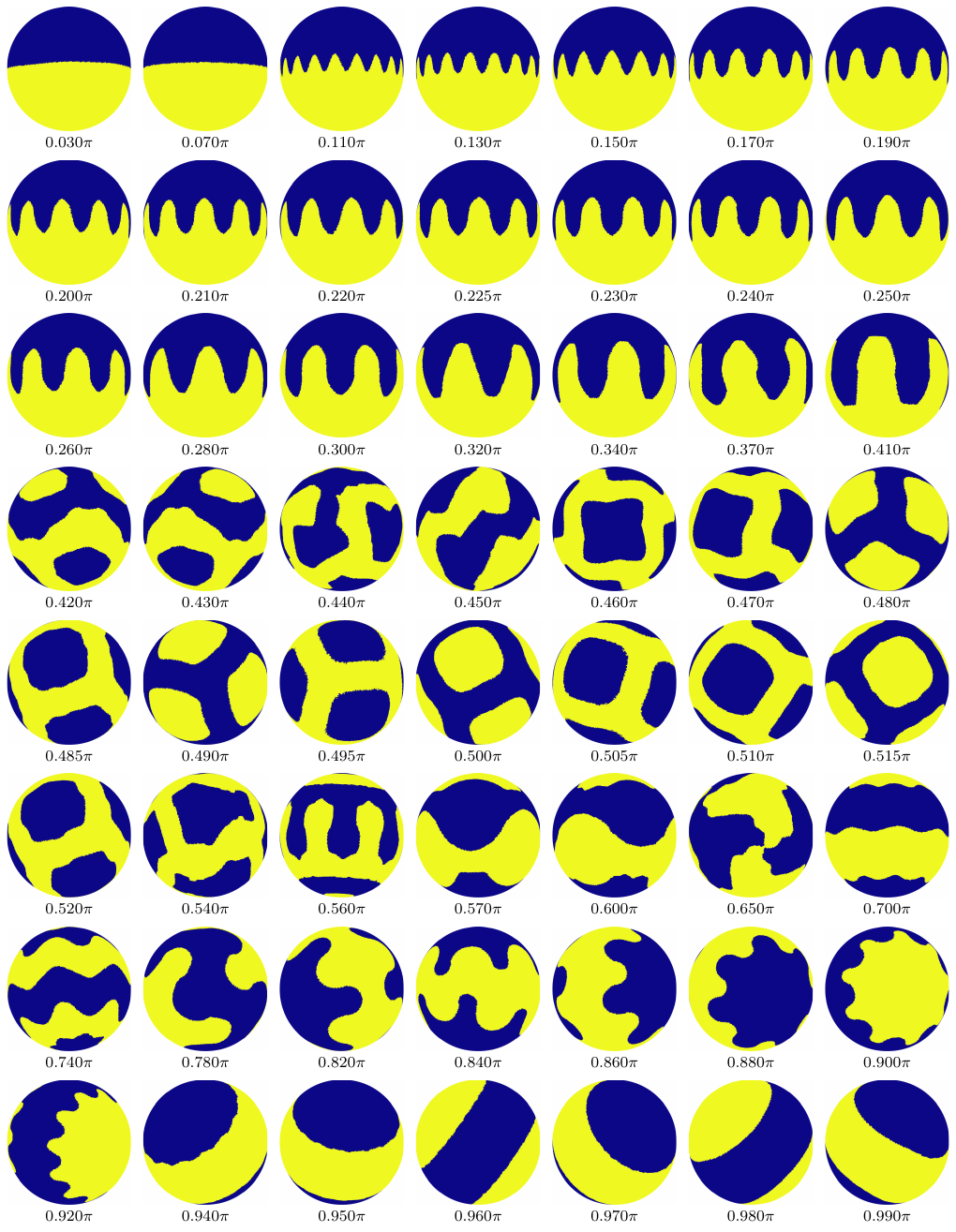}
    \caption{Optimal grasshopper spin configurations in the one-lawn setup without the antipodal constraint for different values of the jump $\theta$.}
    \label{fig:non-antipodal-shapes}
\end{figure*}

The labyrinth-shaped lawns around $\theta=\pi/2$ in the antipodal two-lawn setup look qualitatively similar to the complementary labyrinth-shaped lawns. The two independent lawns are typically offset in this regime as well. But in contrast to the one-lawn setup, there is no stripes regime in the antipodal two-lawn setup. Due to a $\theta\leftrightarrow\pi-\theta$ symmetry of Eq.~\eqref{eq:continuous} the cogwheel regime returns for angles $\theta\gtrsim0.59\pi=\pi-0.41\pi$. The numerically generated lawn shapes are exactly the same, up to inversion of one of the lawns. The corresponding probabilities are also the same and for $\theta=\pi$ the probability equals one.

\subsection{Non-antipodal complementary (one-lawn) setup}
\label{sec:nonantipodalonelawn}
It is interesting to examine how the antipodal condition that arises from considering local hidden variable models for quantum measurements affects the optimal grasshopper shapes. For instance, it restricts the number of cogs in the cogwheel regime (and the number of cog-like perturbations in the stripes regime) to odd numbers and it is also necessary for the special cases at $\theta=\pi/2$ and $\theta=\pi$. A straightforward generalization is to define the problem on the sphere without the antipodal condition, merely demanding that half of the sphere must be covered by the lawn. The results for this setup are presented in this section.  Numerically found optimal shapes are shown in Fig.~\ref{fig:non-antipodal-shapes}.
The complement of an optimal shape is also optimal, since the general grasshopper success probability functional given by Eq.~\eqref{eq:continuous} is symmetric in $L_1$ and $\overline{L}_2$.


As in the other two setups discussed previously, the optimal lawn configurations for jump angles $\theta\lesssim0.41\pi$ are cogwheels. The number of cogs for the first mode is the integer closest to $2\pi/\theta$, as shown in Fig.~\ref{fig:cognumber}. The absence of the antipodal restriction for the number of cogs to be odd implies that the cog number now follows the $2\pi/\theta$ curve more closely. The shapes of the cogs are round; the triangular cog shapes that were observed in the antipodal setup close to even values of $2\pi/\theta$ (where the highest extent of rounding was necessary) are no longer present. Without the antipodal condition, the cogwheel shapes are always favorable compared to the hemispherical lawn and higher modes or mode mixing are not as prevalent, for the same reason.

Beyond the cogwheel regime, the shapes transition to configurations with stripes and islands, which eventually acquire the symmetry of a cube, with six squircles (squares with rounded corners) arranged accordingly that belong to the first lawn and are separated by the second lawn. The configuration is most regular at $\theta=\pi/2$. For larger jump angles the shapes become less regular. Some of these numerically observed patterns, such as combinations of cogs and rings, were also observed in the corresponding antipodal setup. The stripes regime, which arises for $\theta\gtrsim0.57\pi$, as in the antipodal case, also bears some resemblance to the latter, but with significant differences. There are fewer stripes and the cog-like perturbations on the stripes are much more pronounced. Only for jumps $\theta\gtrsim0.95\pi$ are these perturbations no longer perceptible. 

As $\theta$ approaches $\pi$, the lawn forms two identical round caps around the poles, separated by a single stripe. Since the total area of the lawn is $2\pi$, the area of each cap equals $\pi$, corresponding to an angle of $\pi/3$ between the pole and the boundary of each cup. For such a configuration, the grasshopper success probability for $\theta=\pi$ equals one, since in this case each pair of antipodal points belongs to the same lawn. Many other configurations exist fulfilling this condition and thus resulting in the optimal grasshopper success probability, but the two-cap configuration is the one produced by the numerical optimization setup.

\subsection{Comparison of success probabilities}
\label{sec:probs}
\begin{figure*}
    \centering
    \includegraphics[width=\textwidth]{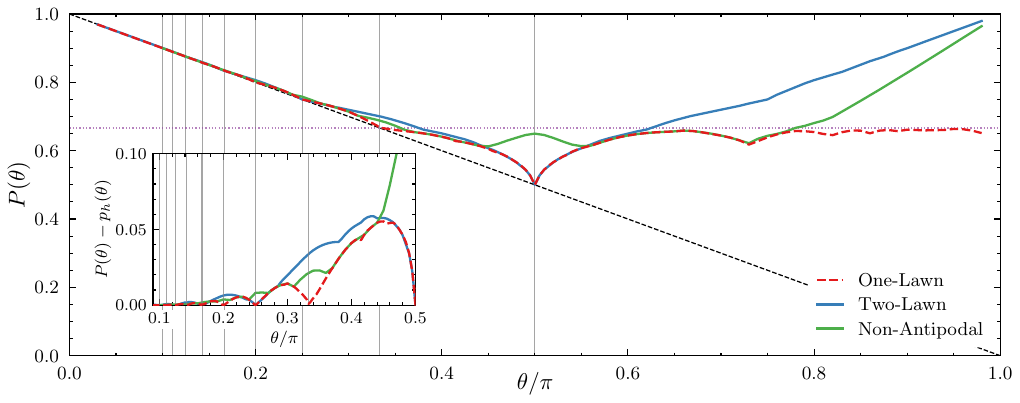}
    \caption{Probabilities of numerically found optimal lawn shapes as function of $\theta$ for the three setups considered: antipodal complementary (one-lawn) setup, antipodal independent (two-lawn) setup, and non-antipodal complementary (one-lawn) setup. The black dashed line is the hemisphere lawn probability $1-\theta/\pi$ and the dotted horizontal line is the analytical limit in the antipodal one-lawn stripes regime, corresponding to $2/3$, see text for discussion. Vertical gray lines denote jump angles of the form $\theta=\theta_q$ for integer $q$. As expected, the antipodal independent lawn probabilities and the non-antipodal complementary lawn probabilities are always greater or equal to the ones in the antipodal complementary setup, since the latter is a more restrictive version of the former two. The antipodal two-lawn probability is symmetric around $\pi/2$. The inset shows the difference between the optimal lawn probabilities and the hemisphere probability for jump angles below $\pi/2$ in all three setups.}
    \label{fig:probability_graph}
\end{figure*}
The optimal lawn probabilities for the different setups discussed above are plotted in Fig.~\ref{fig:probability_graph} and compared to the hemisphere probability. The corresponding comparison to quantum correlations is presented in the parallel publication~\cite{llamas2025sphericalletter}.
We note the agreement with analytical results, where the latter are available. As expected, the antipodal independent lawn probabilities and the non-antipodal complementary lawn probabilities are always greater or equal to the ones in the antipodal complementary setup, since the latter is a more restrictive version of the former two. The probability curve in the antipodal independent setup is symmetric around $\theta=\pi/2$, following the symmetry of the corresponding integral. For angles of the form $\theta=\theta_q$, the optimal grasshopper success probability is equal to the hemispherical result $1-\theta/\pi$ for any integer $q$ in the antipodal one-lawn setup, and for even integers $q$ in the antipodal two-lawn setup, as discussed above. For $\theta=\pi/2$, the grasshopper success probability is $1/2$ in these setups, as expected. When $\theta$ exactly equals $\pi$ the antipodal one-lawn probability is 0 by construction. However, for angles just below the critical value, the probability indeed tends to the limit of $2/3$, which was derived for the stripes regime in Sec.~\ref{sec:antipodalonelawn}.

It is interesting to see where the different grasshopper setups produce the same or nearly the same probabilities. For small jump angles in the cogwheel regime, all probability curves overlap over small intervals between successive values of $\theta_q$. For the antipodal and non-antipodal one-lawn setups this is expected, as in those intervals the lawn configurations are the same. The probability curves for the antipodal one- and two-lawn setups also overlap in the labyrinth regime around $\pi/2$. And in the antipodal and non-antipodal one-lawn setups there is significant additional overlap or near-overlap in the interval $0.55\pi\lesssim\theta\lesssim0.75\pi$. Both setups exhibit similar striped configurations in that regime, although the numbers of cog-like protrusions on the stripes differ.

\subsection{Insights from the spherical harmonics expansion}
\label{sec:harmonics}
Additional insights into the nature of the optimal lawn shapes and the corresponding probabilities can be gained from the spherical harmonics expansion introduced in Sec.~\ref{sec:problemstatement}. In the following, we will focus on the antipodal one-lawn setup, which exhibits the greatest variety of shapes, but a similar analysis can also be performed for the other setups. Equation~\eqref{eq:p-spectral-onelawn} implies that, for fixed $\theta$, the contribution of each spherical harmonic degree to the grasshopper success probability is weighted by $P_\ell(\cos\theta)$. Hence, the optimization challenge is to find a set of coefficients $\{\widehat{\mu}_{\ell m}\}$ that obey the appropriate constraints on $\mu(\mathbf{r})$ outlined in Sec.~\ref{sec:problemstatement}, and that favor those values of $\ell$ where $P_\ell(\cos\theta)$ is positive and largest.

\begin{figure}[t]
  \centering
  \includegraphics[width=\columnwidth]{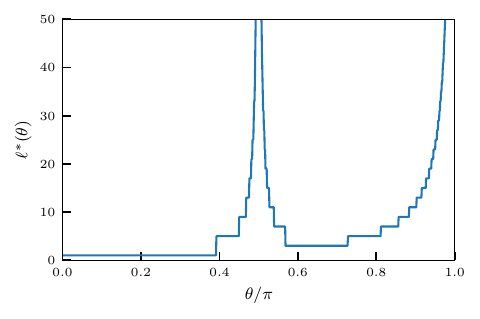}
  \caption{Degree $\ell^*(\theta)$ of the dominant contribution $P_{\ell^*}(\cos\theta)$ to the antipodal one-lawn probability in the spherical harmonics representation.}
  \label{fig:ellstar}
\end{figure}

While we cannot concentrate all of the spectral weight into one dominant $\ell$ contribution due to the restrictions on the coefficients $\{\widehat{\mu}_{\ell m}\}$, considering the optimal $\ell$ gives a useful upper bound on the attainable grasshopper success probability,
\begin{equation}
P(\theta)\leq\frac{1}{2}+\frac{1}{2}\textnormal{max} P_\ell(\cos\theta),
\end{equation}
where $\ell\geq1$ and $\ell$ odd for the antipodal case. Figure~\ref{fig:ellstar} shows the corresponding optimal value of $\ell$ as a function of the jump angle $\theta$. For small jumps, $\ell^*=1$. It increases to $\ell^*=5$ around $\theta\approx0.392\pi$, which coincides very closely with the numerically observed transition from the cogwheel regime to more complex configurations. It keeps increasing sharply, tending towards infinity as $\theta\rightarrow\pi/2$ (labyrinth regime) and then decreases again to $\ell^*=3$ at around $\theta\approx0.568\pi$, very close to where we numerically observe the onset of the stripes regime. As $\theta$ increases further so does $\ell^*$, and it tends towards infinity again as $\theta\rightarrow\pi$.

\begin{figure*}[t]
  \centering
  \includegraphics[width=\textwidth]{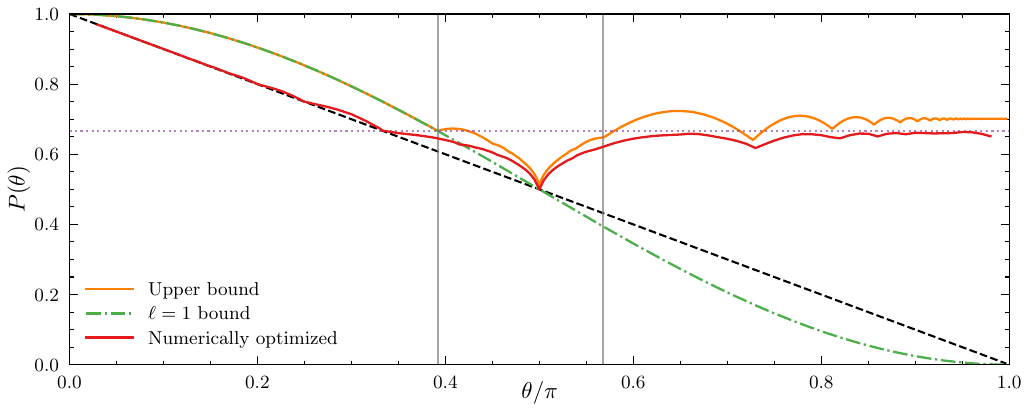}
  \caption{Upper bound $\frac{1}{2}+\frac{1}{2}\textnormal{max} P_\ell(\cos\theta)$ on the grasshopper success probability in the antipodal one-lawn case as function of $\theta$ (orange solid line). For comparison, the probability for the numerically found optimal lawn shape in the same setup is also shown (red solid line). The hemisphere probability (dashed black line) is computed via the spectral sum and agrees with the exact value, providing a consistency check. The green dash-dotted line is the corresponding probability bound when restricting to the $\ell=1$ band, which is dominant in the cogwheel regime. The purple dotted horizontal line is $2/3$ (the analytical limit in the stripes regime) and is shown for reference.}
  \label{fig:probbound}
\end{figure*}

The corresponding upper bound on the probability is plotted in Fig.~\ref{fig:probbound}. The curve has a remarkably similar shape to the actual probability curve obtained numerically, despite being generally above the exact curve, as expected. The figure also shows the probability bound under the restriction $\ell^*=1$, which is the dominant contribution for smaller jumps (in the cogwheel regime). Since $P_1(\cos(\theta))=\cos(\theta)$, this bound equals
\begin{equation}
P(\theta)\leq\frac{1}{2}+\frac{1}{2}\cos(\theta),
\end{equation}
which coincides with the singlet anticorrelation, as discussed in Ref.~\cite{llamas2025sphericalletter}.

The spectral representation allows us to better interpret the qualitative lawn shapes. Cogwheels correspond to dominant sectoral harmonics ($m=\ell$), which exhibit structures near the equator and negligible amplitude outside the tropics \cite{boyd2001book}. Stripes correspond to dominant zonal harmonics ($m=0$), which have the latitudinal structure of ordinary Legendre polynomials.

The spherical harmonics representation of the grasshopper functional also helps to place the observed lawn shapes in the broader context of equilibrium pattern formation. 
It is known that similar patterns can arise across systems with completely different microscopic structures and interactions, including superconductor and magnetic garnet films, ferrofluids, block copolymers, phase-separating mixtures \cite{seul1995patterns,andelman2009patterns} and even epidermal ridges (fingerprints) \cite{Kucken2005fingerprint}. 
The connection between these systems is the presence of modulated phases that arise due to competing interactions that favor structure on different length scales.
This competition often features nonlocal interactions and it can drive symmetry breaking even when the underlying setting is isotropic.
The resulting patterns include stripes and ripples, bubbles and islands, lamellae, and labyrinth-like structures.
Defects such as branching and splitting also commonly appear, especially near crossovers between different patterns.
In the spherical grasshopper system, which also exhibits such patterns, the analogous competition occurs across spherical‑harmonic bands $\ell$, due to the sign-alternating spectral weight $P_\ell(\cos\theta)$ paired with global constraints on the spectral coefficients.
The optimization therefore becomes a constrained redistribution of spectral weight across multiple degrees, which translates into competition across length scales in real space and causes the emergence of stripe-like and ripple-like structures.
Similar patterns also arise in classical Turing-type reaction–diffusion systems, where imposing spherical geometry and curvature can strongly influence which patterns are selected and how defects organize on the surface \cite{varea1999turing,krause2018turing,staddon2024zebrastripes}.

\section{Summary and Conclusions}
\label{sec:conclusions}
In summary, we presented and discussed numerical results for optimal grasshopper lawn shapes in the antipodal one- and two-lawn setups, as well as in the non-antipodal one-lawn setup. We analyzed qualitative and quantitative features of the configurations, including shape, number, and height of cogs in the cogwheel regimes, shape and number of stripes in the stripe regime, etc. When possible, we compared with analytical or semi-analytical calculations. We also juxtaposed the three setups, both in terms of lawn shapes and the corresponding probabilities, and interpreted these in the context of a spherical harmonics expansion.

It is instructive to compare the spherical results to the corresponding solutions of the planar grasshopper problem, which was discussed in detail in Refs.~\cite{goulko2017grasshopper, llamas2023grasshopper}. The planar problem also exhibits cogwheel-shaped optimal solutions for small values of the jump. Since for small jump angles on the sphere the curvature of the domain matters less, it is plausible that the same type of configuration is optimal for the spherical grasshopper problem also. However, the disk-shaped lawn is never optimal in the 2-dimensional planar case, while on the sphere it is an optimal solution for jump angles $\theta_q$. The planar stripes regime also shows some similarity to the spherical stripes regime. On the plane, stripes were observed for jump lengths $d\gtrsim0.88$ (assuming that the planar lawn has area equal to one) and these stripes also exhibit arc-like modulations on the boundaries.

We have also numerically established the nature of the optimal lawn shapes across all $0\leq\theta\leq\pi$ for the one-lawn setup for
general (not necessarily antipodal) lawns.   These do not define local hidden variable models for projective measurements, but are 
mathematically interesting and give further insights about the solutions for the antipodal lawn setup, and in particular into the 
range of $\theta$ for which the antipodal constraint is significant.   
The numerical methods used to study the spherical grasshopper problem are directly applicable also to other problems in geometric combinatorics and can help test the validity of conjectures or provide information that assists the formulation of analytical proofs.
A relevant example for an optimization problem on the sphere is the Double Cap Conjecture \cite{Witsenhausen1974doublecap, Kalai2009doublecap, kalai2015doublecap}, which posits that the largest measure of a Lebesgue measurable subset of the ($n$-dimensional) unit sphere containing no pair of orthogonal vectors is attained by two open caps of geodesic radius $ \pi/4 $ around the north and south poles. While certain bounds on the measure are known \cite{decorte2015doublecap,decorte2022doublecap,bekker2025doublecap}, the maximal measure and the corresponding shapes remain an open problem in all dimensions and are subject to ongoing work. 

We present more detailed numerical solutions for the spherical grasshopper problem and compare them with the corresponding quantum correlations in a parallel publication \cite{llamas2025sphericalletter}, in which we compare the grasshopper success probabilities for the single-lawn and two independent lawn set-ups with the corresponding quantum singlet probabilities.   This identifies the maximal size of the gaps between classical and quantum correlations and the optimally efficient non-locality tests involving Bell inequalities for randomly chosen axes separated by a fixed angle.

One might hope that generalizations of these Bell inequalities, involving two or more possible angular axis separations, would be more efficient still, since the adversary has even less information to exploit and the optimal lawn configurations for different angles are generally different. Indeed, the methods we describe below can be extended to analyze these more general tests.  While such a project is beyond the scope of this paper, our results suggest that even more efficient two- or multi-angle Bell inequalities could indeed be found.

Similarly interesting and practically relevant questions also arise in studying tests of bipartite entanglement and Bell non-locality for non-maximally entangled states of qubits, for general qudit states, and for multi-partite states.   
In a different direction, it would be good to understand more precisely the general interplay between entanglement, classical communication and Bell non-locality, illustrated for example by the results of Bacon-Toner \cite{PhysRevLett.91.187904} and Brassard et al. \cite{PhysRevLett.83.1874}.   One can see these results as showing how much classical communication is required to eliminate the gap between quantum and Bell-local hidden variable correlations for specific tests on specific states.   Again, our methods offer a new approach to more general questions of this type.   

In previous work \cite{llamas2023grasshopper} we pointed out potential connections between the planar grasshopper Ising model and trapped dipolar Bose-Einstein condensates, which also have long-range interactions and in certain regimes exhibit similar droplet patterns to some of the optimal grasshopper shapes \cite{hertkorn2021dipolar, schmidt2021dipolar}. In addition to cogwheel and stripes regimes, these include labyrinth-like shapes, which were not observed in the planar grasshopper system, but which do emerge in the spherical model.   It will be interesting to understand better the classes of long-range statistical models and their relationships, whether the structures we find in the grasshopper model are typical, and indeed whether they might be signals of a form of universality in long-range models.

\section*{Acknowledgements}
We thank Carlo Piovesan for generously sharing unpublished numerical results including some of the lawn types
described in this work.
Some of the results in this paper have been derived using the healpy and HEALPix numerical packages.
Portions of the spherical harmonics exposition benefited from suggestions from ChatGPT (OpenAI). All statements were independently drafted and verified by the authors.

This work is supported by the NSF under Grant No. PHY-2112738 and Grant No. OSI-2328774 (OG and DL). OG also acknowledges support under NSF Grant No. PHY-2441282.
The work of DL was supported in part by College of Science and Mathematics Dean's Doctoral Research Fellowship through fellowship support from Oracle, project ID R0000000025727.
AK acknowledges financial support from project OPP640, funded by the Science and Technology Facilities Council's International Science Partnerships Fund.
AK was supported in part by Perimeter Institute for
Theoretical Physics. Research at Perimeter Institute is supported by
the Government of Canada through the Department of Innovation, Science
and Economic Development and by the Province
of Ontario through the Ministry of Research, Innovation and Science. 

\appendix
\section{Direct numerical calculation of grasshopper success probabilities for triangular cogs}
\label{sec:appendix}
Here is an outline of the procedure we use for $k$ cogs with boundaries that are spherical triangles, i.e., the boundaries are composed of segments of great circles. We denote the northern hemisphere by $H$ and the set of $k$ ``teeth", the intersection of the cogs with the southern hemisphere, by $\cal{T}$. The antipodal image of any set $A$ is denoted by $\overline{A}$. The lawn $L$ under consideration can be expressed as $H$ together with the set $\cal{T}$ added and the set $\overline{\cal{T}}$ (``gaps") subtracted, i.e., $L=H+\cal{T}-\overline{\cal{T}}$. We want to compute the probability of a jump from the uniform distribution over $L$ onto $L$. Since $L$ has area $2\pi$, in our application any part of $L$ has initial probability density~$\frac1{2\pi}$. 

\begin{figure}[t]
\includegraphics[width=\columnwidth]{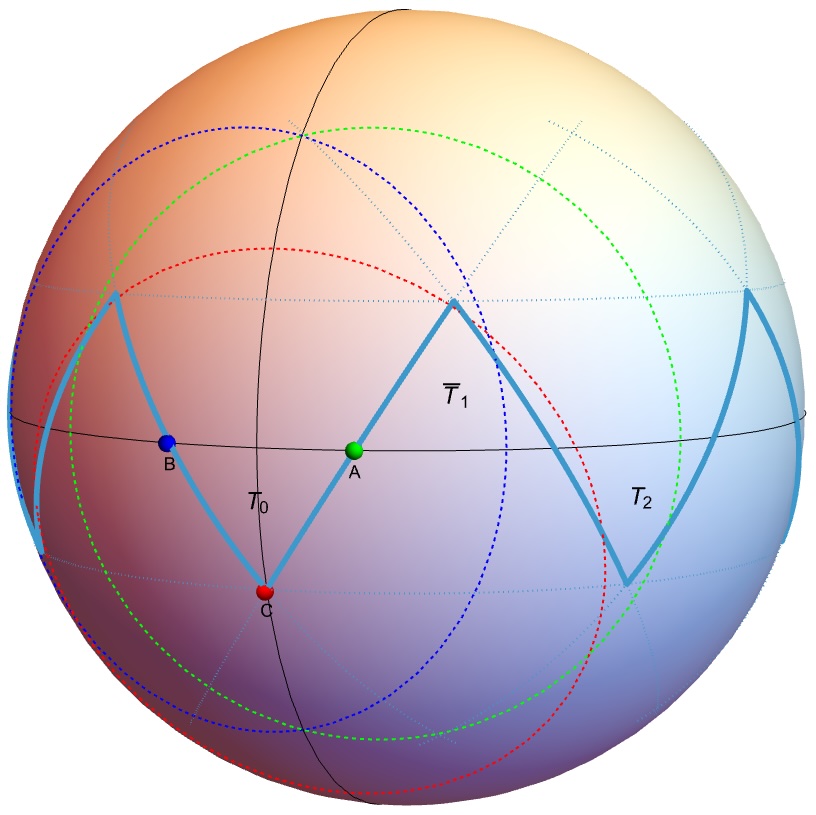}
\caption{Illustration of the direct probability computation for triangular cogged lawns for $\theta=\pi/4$ and $7$ cogs with height equal to $0.332$. The lawn is centered at the top (the north pole) and bounded by the thick blue line. Teeth $T_0$ and $T_2$ are labeled, as is the gap ${\overline T_1}$. The jump circles of radius $\pi/4$ centered on the three vertices of $T_0$ are shown with dashed lines of appropriate colors.}
\label{fig:triangularconstruct}
\end{figure}

We use the following notation.
For sets $A,B$ we denote the jump probability from $A$ to $B$ simply by juxtaposition or product. For example, our aim is to compute $L^2$. 
We observe that for any sets $A,B$, (i) $AB = BA$, and (ii) $AB=\overline{A}\,
\overline{B}$.  So
\begin{align*}
&L^2 = (H+{\cal T}-\overline{\cal T})^2 \\
  &= H^2 + H{\cal T} + {\cal T}H - H\overline{\cal T} - \overline{\cal T}H + {\cal T}^2 + \overline{\cal T}^2 -  \cal{T}\overline{\cal T} - \overline{\cal T}{\cal T}\\
  &= H^2 + 2{\cal T}H - 2{\cal T}\overline{H}  + 2{\cal T}^2  - 2{\cal T}\overline{\cal T}.
\end{align*}

Now, ${\cal T}$ consists of $k$ teeth and $\overline{\cal T}$ consists of $k$ gaps (antipodal images of teeth), and we label the alternating sequence of teeth and gaps around the equator as $T_0,{\overline T}_1,T_2,{\overline T}_3,\dots,T_{2k-2},{\overline T}_{2k-1}$.

Our computational task is illustrated in Fig.~\ref{fig:triangularconstruct} for the case of jump distance $\pi/4$ with $7$ triangular cogs of height $0.332$. 
It can be seen that for this choice of parameters there can be no jump from $T_0$ to itself nor from $T_0$ to any teeth or gaps which are not those nearest to $T_0$. The same holds true for the other parameter combinations under consideration. It will therefore be sufficient for us to calculate the jump probabilities from $T_0$ to ${\overline T}_1$, $T_2$, $H$ and to ${\overline H}$. For the latter, we observe that, since $H$ and ${\overline H}$ are disjoint and cover the whole sphere, for any point $p$ the probability $p{\overline H} = 1- pH$. 
From symmetry we see that 
$${\cal T}^2 = 2k T_0T_2 \quad \mathrm{ and } \quad {\cal T}\overline{\cal T} = 2k T_0{\overline T_1} .$$

The jump circle about any point $p$ on the sphere and any great circle have at most two intersections, which we can find as explicit solutions of a trigonometric equation. For our example in Fig.~\ref{fig:triangularconstruct}, the red jump circle about the point $C$ has solutions in anticlockwise order, beginning in the south, entering $T_2$ from the left and leaving at the equator, entering ${\overline T_1}$ from the right and leaving on the left. 
Calculating the lengths of these two arcs gives us the point $C$'s contribution to the probabilities $T_0T_2$ and $T_0{\overline T_1}$ respectively. We can do a numerical integration over the whole of $T_0$ to get these probabilities with high accuracy. Calculation of the probability $T_0H$, and hence $T_0{\overline H}$, is similar but simpler.  For other cog heights, jump distances and numbers of cogs, we might need the computation of probabilities for jumps from $T_0$ to other teeth or gaps.

In general, to derive the success probabilities from an arbitrary point $p$ in $T_0$, we calculate all the intersections of the jump circle about $p$ with the equator and with all other great circles defining the boundaries of teeth and gaps of interest, sort them by their angle about $p$ and hence derive the arc lengths of this jump circle that lie within each such tooth or gap. 

\section*{References}
\bibliography{grasshopper,postdocbiblio}

\end{document}